\documentclass[prd,nofootinbib,preprint,superscriptaddress]{revtex4}

\usepackage{amsmath, amssymb, amsthm, graphicx, epsfig, fancyhdr,epsfig, slashed, mathrsfs}
\usepackage[T1]{fontenc}
\usepackage{tikzsymbols}
\usepackage{natbib}
\usepackage{float}
\usepackage[thinc]{esdiff}
\usepackage{feynmf}
\usepackage{tikzsymbols}
\usepackage{natbib}
\usepackage{ amssymb }
\usepackage{amsmath}
\usepackage{dsfont}
\usepackage{cancel}
\usepackage[titletoc]{appendix}
\usepackage{float}

\usepackage{tikz,xcolor,hyperref}

\usepackage{slashed}

\definecolor{lime}{HTML}{A6CE39}
\DeclareRobustCommand{\orcidicon}{
	\begin{tikzpicture}
	\draw[lime, fill=lime] (0,0) 
	circle [radius=0.2] 
	node[white] {{\fontfamily{qag}\selectfont \tiny ID}};
	\draw[white, fill=white] (-0.0625,0.095) 
	circle [radius=0.007];
	\end{tikzpicture}
	\hspace{-2mm}
}

\foreach \x in {A, ..., Z}{\expandafter\xdef\csname orcid\x\endcsname{\noexpand\href{https://orcid.org/\csname orcidauthor\x\endcsname}
			{\noexpand\orcidicon}}
}

\usepackage{tikz,xcolor,hyperref}

\definecolor{lime}{HTML}{A6CE39}
\DeclareRobustCommand{\orcidicon}{
	\begin{tikzpicture}
	\draw[lime, fill=lime] (0,0) 
	circle [radius=0.2] 
	node[white] {{\fontfamily{qag}\selectfont \tiny ID}};
	\draw[white, fill=white] (-0.0625,0.095) 
	circle [radius=0.007];
	\end{tikzpicture}
	\hspace{-2mm}
}

\foreach \x in {A, ..., Z}{\expandafter\xdef\csname orcid\x\endcsname{\noexpand\href{https://orcid.org/\csname orcidauthor\x\endcsname}
			{\noexpand\orcidicon}}
}

\newcommand{\be}{\begin{equation}}
\newcommand{\ee}{\end{equation}}
\newcommand{\bea}{\begin{eqnarray}}
\newcommand{\eea}{\end{eqnarray}}

\usepackage{amsmath}
\usepackage{braket}
\usepackage{graphicx}
\usepackage{mathrsfs}
\usepackage{xcolor}
\usepackage{float}
\usepackage{gensymb}
\usepackage{amssymb}
\usepackage{slashed}
\usepackage{latexsym}
\usepackage{breqn}

\newcommand{\ba}{\begin{eqnarray}}
\newcommand{\ea}{\end{eqnarray}}
\newcommand{\bi}{\begin{itemize}}
\newcommand{\ei}{\end{itemize}}

\newcommand{\x}{\star}

\newcommand{\arXivold}[2]{\href{http://arxiv.org/pdf/#1}{{\tt #2/#1}}}

\begin{document}
\title{Corrected Calculation for the Non-local Solution to the $g-2$ Anomaly and Novel Results in Non-local QED}
\author{Fayez Abu-Ajamieh}
\email{fayezajamieh@iisc.ac.in}
\affiliation{Centre for High Energy Physics; Indian Institute of Science; Bangalore; India}

\author{Nobuchika Okada}
\email{okadan@ua.edu}
\affiliation{
Department of Physics and Astronomy; 
University of Alabama; Tuscaloosa; Alabama 35487; USA}

\author{Sudhir K. Vempati}
\email{vempati@iisc.ac.in}
\affiliation{Centre for High Energy Physics; Indian Institute of Science; Bangalore; India}

\begin{abstract}
We provide the corrected calculation of the $(g-2)_{\mu}$ in non-local QED previously done in the literature. In specific, we show the proper technique for calculating loops in non-local QED and use it to find the form factors $F_{1}(q^{2})$ and $F_{2}(q^{2})$ in non-local QED. We also utilize this technique to calculate some novel results in non-local QED, including calculating the correction to the photon self-energy, the modification to the classical Coulomb potential, the modification to the energy levels of the hydrogen atom, and the contribution to the Lamb shift. We also discuss charge dequantization through non-locality, and show that the experimental bounds on the electric charge on Dirac neutrinos, translate into strong flavor-dependent bounds on the scale on non-locality that range between $10^{5} - 10^{10}$ TeV. We also discuss the inconsistencies of unrenormalized non-local Quantum Field Theories (QFTs) and the need for renormalizing them, even when they are free from UV divergences.
\end{abstract}
\maketitle

\section{Introduction}\label{sec:I}
String-inspired non-local QFTs \cite{Witten:1985cc, Kostelecky:1988ta, Kostelecky:1989nt, Freund:1987kt, Freund:1987ck, Brekke:1987ptq, Frampton:1988kr, Tseytlin:1995uq, Seiberg:1999vs, Siegel:2003vt,Biswas:2004qu, Calcagni:2013eua, Calcagni:2014vxa} have attracted significant attention in the recent years, due to their apparent ability to solve several problems in high-energy physics. The fact that non-local QFTs are free from UV divergences and do not introduce any new poles to amplitudes made them suitable for furnishing solutions to several problems, such as the hierarchy problem \cite{Abu-Ajamieh:2023syy, Krasnikov:1987yj, Moffat:1988zt, Moffat:1990jj} and finding bouncing solutions to gravity \cite{Biswas:2005qr}. The good behavior of string-inspired non-local QFTs with infinite derivatives, arises from the non-locality form factor that modulates the kinetic term, usually taken as the exponential of an entire function of the the d'Alembertian operator,
\begin{equation}\label{eq:NL_form factor}
S_{\text{NL}} = \int d^{4}x\Big[ -\frac{1}{2} \phi e^{\frac{\Box +m^{2}}{\Lambda^{2}}}(\Box + m^{2}) \phi - V(\phi) \Big],
\end{equation}
where $\Lambda$ is the scale of non-locality, i.e. the scale at which non-local effects come into play. In particular, it is easy to see that loop amplitudes become modulated by a factor $\sim e^{-s/\Lambda^{2}}$, thereby making  UV divergences exponentially suppressed for energy scales larger than $\Lambda$. This renders non-local QFTs formulated in this manner super-renormalizable in principle. Another attractive feature of this formulation, lies in the fact that an exponential form factor does not introduce any new poles in the propagators, and therefore does not introduce any new degrees of freedom compared with the local version of the theory.

In this paper, we will be concerned mainly with non-local QED. The first realistic formulation of non-local QED was presented in \cite{Biswas:2014yia}. The non-local QED Lagrangian in the Feynman gauge  can be expressed as follows
\begin{equation}\label{eq:NL_QED_Lag}
\mathcal{L} = -\frac{1}{4}F_{\mu\nu}e^{\frac{\Box}{\Lambda_{g}^{2}}}F^{\mu\nu} + \frac{1}{2}\Big[i \overline{\Psi}e^{-\frac{\nabla^{2}}{\Lambda_{f}^{2}}}(\slashed{\nabla}+m)\Psi +h.c. \Big],
\end{equation}
where the covariant derivative is given by
\begin{equation}\label{eq:Covariant_der}
\nabla_{\mu} = \partial_{\mu} + i Q e A_{\mu} \hspace{5mm} \implies \hspace{5mm} \nabla^{2} = \Box + iQ e (\partial \cdot A + A \cdot \partial) - Q^{2} e^{2} A^{2}.
\end{equation}

Notice that while the ordinary derivative is used in the gauge sector, the covariant derivative has to be used in the fermion sector to make the Lagrangian gauge invariant. Also notice that in general, the non-locality scales for the gauge and fermion sectors needn't be the same. In fact, as we will see later on, non-locality (at least in the fermion sector), should be flavor-dependent. The Feynman rules corresponding to the propagators and interaction vertex are given by
\begin{eqnarray}
\Pi_{g}^{\mu\nu}(p) & = & \frac{-i g^{\mu\nu}e^{\frac{p^{2}}{\Lambda_{g}^{2}}}}{p^{2}+ i\epsilon}, \label{eq:photon_propagator} \\
\Pi_{f}(p) & = & \frac{ie^{\frac{p^{2}}{\Lambda_{f}^{2}}}(\slashed{p}+m)}{p^{2}-m^{2}+i\epsilon},  \label{eq:fermion_propagator}\\
V^{\mu}(q_{1},q_{2}) & = & i\frac{Q e}{2}\Bigg[ (q_{1}^{\mu}\slashed{q}_{2}+q_{2}^{\mu}\slashed{q}_{1})\Bigg(\frac{e^{\frac{q_{1}^{2}}{\Lambda_{f}^{2}}}-e^{\frac{q_{2}^{2}}{\Lambda_{f}^{2}}}}{q_{1}^{2}-q_{2}^{2}} \Bigg) + \Big( e^{\frac{q_{1}^{2}}{\Lambda_{f}^{2}}}+e^{\frac{q_{2}^{2}}{\Lambda_{f}^{2}}}\Big)\gamma^{\mu}\Bigg], \label{eq:vertex}
\end{eqnarray}
where $q_{1,2}$ are the momenta of the (outgoing) fermions. The details for deriving these rules are provided in \cite{Biswas:2014yia}, where we refer the interested reader.

Recently, there has been an attempt to solve the $(g-2)_{\mu}$ anomaly \cite{Capolupo:2022awe} through utilizing the above non-local QED extension. As it is well-known, there is a discrepancy between the theoretically predicted (see \cite{Aoyama:2020ynm} and the references therein) and experimentally measured \cite{Muong-2:2006rrc,Muong-2:2021ojo, Muong-2:2021ovs,Muong-2:2021vma} magnetic dipole moment of the muon, which currently stands at a significance of $4.2\sigma$\footnote{The recent results from Fermilab \cite{Muong-2:2023cdq} brings the world average of the discrepancy to $249(59) \times 10^{-11}$, which brings its significance to the $5.0 \sigma$ threshold. However, we should point out that recent high-precision QCD lattice simulations \cite{Borsanyi:2020mff, Ce:2022kxy, ExtendedTwistedMass:2022jpw} appear to agree with the experimental measurements, thus reducing the anomaly and placing tension between the lattice approach and the data-driven approach. We will ignore this tension here as it is irrelevant for our discussion.},
\begin{equation}\label{eq:anomaly}
\Delta a_{\mu} = a_{\mu}^{\text{Exp}} - a_{\mu}^{\text{SM}} = 251(59) \times 10^{-11}.
\end{equation}

Other recent proposals to explain the anomaly include utilizing the SM Effective Field Theory (SMEFT), \cite{Abu-Ajamieh:2022nmt, Abu-Ajamieh:2023qvh, Buttazzo:2020ibd, Yin:2020afe, Fajfer:2021cxa, Aebischer:2021uvt, Allwicher:2021jkr, Cheung:2021iev}, to which other proposal, such as non-locality, can be mapped. The authors in \cite{Capolupo:2022awe} used the non-local QED extension above, in order to calculate the QED form factors $F_{1}(q^{2})$ and $F_{2}(q^{2})$, and use them to explain the $(g-2)_{\mu}$ discrepancy by suitably setting the scale of non-locality to generate the anomaly, where the authors found that a scale on non-locality for the muon of $\Lambda_{\mu} = 4.384$ TeV would be sufficient to generate the anomaly. 

While the argument in \cite{Capolupo:2022awe} is in general correct, the authors nonetheless made errors in their calculation. In specific, the authors calculated the correction to the form factors at tree-level and at one-loop level, and while their tree-level calculation is correct, their calculation at 1-loop is erroneous. In addition, there is another conceptual error related to interpreting the results. The authors claim that one way to interpret their results is to assume violation of gauge invariance. We show that this is incorrect, as gauge invariance will always be preserved.

In this paper, we fix the 1-loop calculation made in \cite{Capolupo:2022awe} by recalculating the form factors at 1-loop using the correct treatment for the non-local form factor, and we provide the correct interpretation of the results. In addition, we utilize our machinery to obtain other 1-loop results in non-local QED, such as the photon self-energy, the modification to the electric potential, and the contribution to the Lamb shift. We also discuss charge dequantization that corresponds to non-locality, and we utilize it to set stronger bounds on the scale of non-locality from the bounds on millicharged particles.

We point out that after finalizing this paper, we found that some of the authors of \cite{Capolupo:2022awe} extended their  treatment in \cite{Capolupo:2023kuu} to calculate several phenomenological results in non-local QED, some of which overlap with our results. This includes calculating the non-local corrections to the photon propagator, the Coulomb potential, the Lamb shift, the electrostatic force, and the running of the QED coupling $\alpha$ in non-local QED. However, the calculation in \cite{Capolupo:2023kuu} utilized the same erroneous treatment used in \cite{Capolupo:2022awe} (see \ref{sec:III} for the discussion of this erroneous treatment). As all their results are predicated upon the erroneous loop correction to the photon propagator, all these results are also incorrect. In addition, they claim that non-locality already arises in the photon propagator at the tree-level, which is incorrect, as according to \cite{Briscese:2015zfa}, any non-interacting non-local QFT is in fact local, which can be shown through a simple change of variable of the wavefunction that absorbs the non-locality on the non-interacting QFT in the definition of the field itself. Therefore, All the results in \cite{Capolupo:2023kuu} are erroneous.

This paper is organized as follows: In Section \ref{sec:II}, we review the results obtained in \cite{Capolupo:2022awe} and highlight the errors they committed, then we provided the corrected calculation of the form factors in Section \ref{sec:III}. In Section \ref{sec:IV}, we present our additional novel results in non-ocal QED at 1-loop. We also comment on the formalism of non-local QED with infinite derivatives based on our result. Finally, we conclude in Section \ref{sec:V}.

\section{Review of the Previous Non-local $(g-2)_{\mu}$ Results}\label{sec:II}
\begin{figure}[!t]
\centerline{\begin{minipage}{0.75\textwidth}
\centerline{\includegraphics[width=300pt]{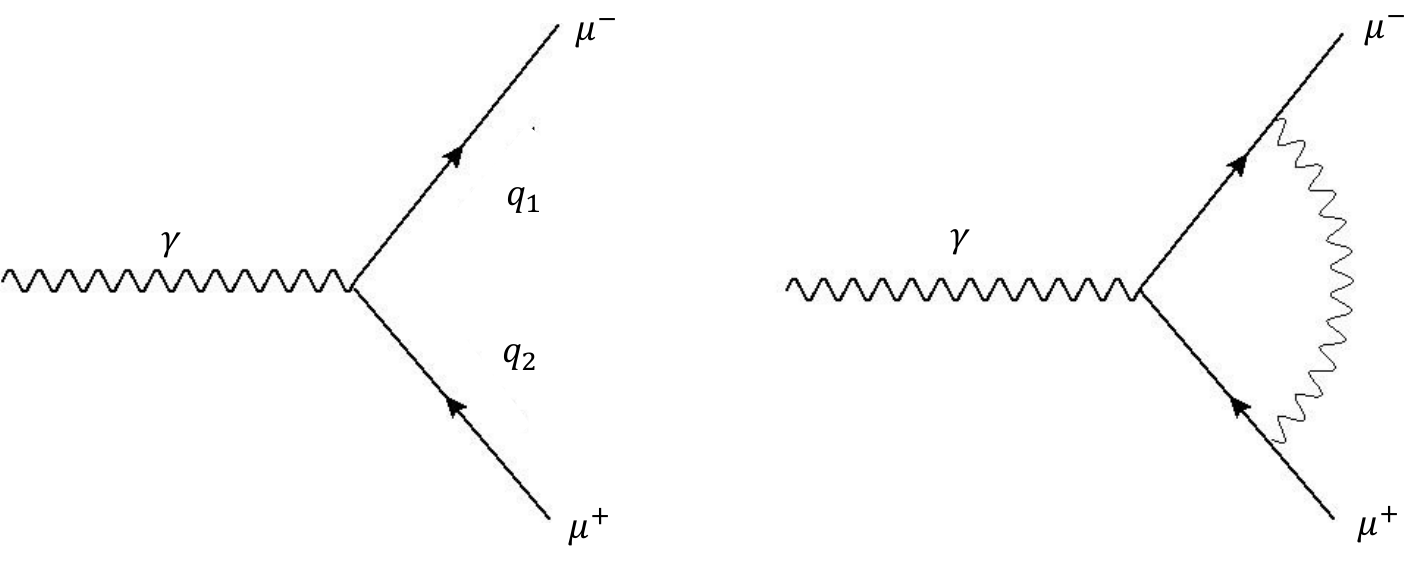}}
\caption{\small Tree-level (left) and 1-loop (right) contributions to the $(g-2)_{\mu}$ anomaly. The propagators and vertices are non-local.}
\label{fig1}
\end{minipage}}
\end{figure}
In this section, we quickly review the main results in \cite{Capolupo:2022awe} and highlight the errors committed by the authors. The non-local contribution to the 	$(g-2)_{\mu}$ was evaluated by calculating the tree-level and the 1-loop corrections to the $\gamma\mu^{+}\mu^{-}$ vertex shown in Figure. (\ref{fig1}). At tree-level, the matrix element can be written as
\begin{equation}\label{eq:Tree_mat1}
	i\mathcal{M}^{\mu} = ie \overline{u}(q_{2})\Gamma^{\mu}(p) u(q_{1}),
\end{equation}
where $p = q_{2}-q_{1}$, and the vertex function $\Gamma^{\mu}$ is expressed in terms of the form factors $F_{1}$ and $F_{2}$ as follow
\begin{equation}\label{eq:VertexFormFactors}
\Gamma^{\mu}(p) =  F_{1}(p^{2})\gamma^{\mu} +F_{2}(p^{2})\frac{i \sigma^{\mu\nu}}{2m}p_{\nu}.
\end{equation}

In non-local QED, $\Gamma^{\mu}(p)$ is given by eq. (\ref{eq:vertex}). The Dirac equation and the Gordon identity can be used to bring the matrix element in eq. (\ref{eq:Tree_mat1}) to the form
\begin{equation}\label{eq:Tree_mat2}
\mathcal{M}^{\mu} = e \overline{u}(q_{2})  \Bigg\{ \Bigg[ m^{2} \Bigg( \frac{e^{\frac{q_{2}^{2}}{\Lambda_{f}^{2}}}-e^{\frac{q_{1}^{2}}{\Lambda_{f}^{2}}}}{q_{2}^{2}-q_{1}^{2}}\Bigg) +\frac{1}{2} \Big( e^{\frac{q_{2}^{2}}{\Lambda_{f}^{2}}}+ e^{\frac{q_{1}^{2}}{\Lambda_{f}^{2}}}\Big) \Bigg]\gamma^{\mu} -\frac{imp_{\nu}}{2}\sigma^{\mu\nu}\Bigg( \frac{e^{\frac{q_{2}^{2}}{\Lambda_{f}^{2}}}-e^{\frac{q_{1}^{2}}{\Lambda_{f}^{2}}}}{q_{2}^{2}-q_{1}^{2}}\Bigg) \Bigg\} u(q_{1}),
\end{equation}
and for the muon we have $\Lambda_{f} \equiv \Lambda_{\mu}$. Comparing (\ref{eq:Tree_mat2}) with (\ref{eq:Tree_mat1}) and (\ref{eq:VertexFormFactors}), we can immediately identify
\begin{eqnarray}
F_{1}(p^{2}) & = &  m^{2} \Bigg( \frac{e^{\frac{q_{2}^{2}}{\Lambda_{\mu}^{2}}}-e^{\frac{q_{1}^{2}}{\Lambda_{\mu}^{2}}}}{q_{2}^{2}-q_{1}^{2}}\Bigg) +\frac{1}{2} \Big( e^{\frac{q_{2}^{2}}{\Lambda_{\mu}^{2}}}+ e^{\frac{q_{1}^{2}}{\Lambda_{\mu}^{2}}}\Big), \label{eq:F1_tree}\\
F_{2}(p^{2}) & = &  -m^{2} \Bigg( \frac{e^{\frac{q_{2}^{2}}{\Lambda_{\mu}^{2}}}-e^{\frac{q_{1}^{2}}{\Lambda_{\mu}^{2}}}}{q_{2}^{2}-q_{1}^{2}}\Bigg). \label{eq:F2_tree}
\end{eqnarray}

Setting $\mu^{\pm}$ to be on-shell, i.e. $q_{1}^{2} = q_{2}^{2} = m_{\mu}^{2}$, we see that the muon $g$ factor is given by
\begin{equation}\label{eq:g_factor}
g_{\mu} = 2\Big( F_{1}(p^{2}=0) +  F_{2}(p^{2}=0)\Big)  = 2 e^{\frac{m_{\mu}^{2}}{\Lambda_{\mu}^{2}}}, \hspace{5mm} \implies (g-2)_{\mu} = 2(e^{\frac{m_{\mu}^{2}}{\Lambda_{\mu}^{2}}} - 1) \simeq 2\frac{m_{\mu}^{2}}{\Lambda_{\mu}^{2}},
\end{equation}
and therefore, generating the anomaly in eq. (\ref{eq:anomaly}) at tree-level only requires $\Lambda_{\mu} \simeq 3$ TeV.\footnote{The value of $\Lambda_{\mu} \simeq 4.384$ TeV cited in \cite{Capolupo:2022awe} includes the (incorrect) loop correction.} This is the main result in \cite{Capolupo:2022awe} and it is indeed correct.

Matters are different, however, when considering the 1-loop contribution shown on the RHS of Figure \ref{fig1}. When calculating the 1-loop diagram, the authors in \cite{Capolupo:2022awe} expanded the non-locality form factor as follows:
\begin{equation}\label{eq:expansion}
	e^{\pm s/\Lambda_{f}^{2}} = 1 \pm \frac{l^{2}+l\cdot(\cdots) + p_{ext}^{2}}{\Lambda_{f}^{2}} + \cdots,
\end{equation}
where $ p_{ext}$ are the external momenta and $l$ is the internal momentum in the loop to be integrated over. The authors claimed that all higher-order corrections were divergent and needed regularization, keeping only the non-divergent leading order for their calculation, and keeping the non-locality form factor with only one vertex/propagator at a time in their calculation, while assuming the other vertices/propagators to be local in the matrix element. This way, they furnished six sub-diagrams for the 1-loop calculation by assigning the non-locality to a different vertex/ propagator each time, then they summed all six matrix elements together. This treatment is erroneous, as expanding in powers of $l^{2}/\Lambda_{f}^{2}$ breaks down when $l \gtrsim \Lambda_{f}$, which is bound to happen when the integral over $l$ is carried out. Thus, all the results in \cite{Capolupo:2022awe} at 1-loop are incorrect.

Contrary to the claim of the authors in \cite{Capolupo:2022awe}, the momentum integrals in non-local QFTs are actually non-divergent, although they might still need regularization in some cases to keep the threshold corrections within acceptable limits as discussed in \cite{Abu-Ajamieh:2023syy} and as we show later on. In fact, the main motivation for introducing an exponential non-locality form factor was to render the theory super-renormalizable and free from any UV divergences. Expanding the non-locality form factor as in eq. (\ref{eq:expansion}) while truncating the said expansion at a certain order in the internal momentum $l$ invalidates this renormalizability, as now the infinite series in $l$ (which sums up to a finite exponential), is turned into a finite sum that is now divergent. This is clearly an incorrect prescription to deal with the form of non-locality at loop-level.

As we shall show in the next section, the proper prescription is to expand in powers of the \textit{external momenta} (divided by the scale of non-locality), rather than the loop momentum, i.e. expand in powers of $p_{ext}/\Lambda$. This is because for all relevant observables, including the contribution to the magnetic dipole moment, the external momenta are always small compared to the scale of non-locality, which according to \cite{Biswas:2014yia} is constrained by the LHC data to be larger than $\sim 2.5 - 3$ TeV. Of course, for calculations where the external momenta are large and comparable to the (presumed) scale of non-locality, the expansion in $p_{\text{ext}}/\Lambda$ is not valid, however, when this occurs, the entire Lagrangian in eq. (\ref{eq:NL_QED_Lag}) is no longer valid as an Effective Field Theory (EFT), and must be replaced with the full (and yet unknown) UV theory. This case is not relevant at the low energies of interest for our purposes and we ignore it here. 

\section{Corrected Results for $F_{1}(p^{2})$ and $F_{2}(p^{2})$}\label{sec:III}
In this section, we show how to calculate the matrix element that corresponds to the 1-loop diagram shown on the RHS of Figure \ref{fig1} in non-local QED. Specifically, we show how to extract the form factors $F_{1}(p^{2})$ and $F_{2}(p^{2})$. In our calculation, we limit ourselves to the LO in the expansion of $p_{\text{ext}}/\Lambda$. The matrix element can be written as
\begin{dmath}\label{eq:NLO_matrix}
i\mathcal{M}^{\mu} = \Big(\frac{-ie}{2}\Big)^{3}\overline{u}(q_{2})\int \frac{d^{4}k}{(2\pi)^{4}}V^{\nu}(q_{2},k+q_{2})\Pi_{f}(k+q_{2})V^{\mu}(k+q_{1},k+q_{2})\\
\times \Pi_{f}(k+q_{1})V^{\rho}(q_{1},k+q_{1})\Pi_{g,\nu\rho}(k)u(q_{1}),
\end{dmath}
where $\Pi_{f}$, $\Pi_{g}$ and $V$ are given eqs. (\ref{eq:photon_propagator}),  (\ref{eq:fermion_propagator}) and (\ref{eq:vertex}), respectively. 
\subsection{The Proper Momentum Expansion}\label{sec:III_1}
Consider a non-locality factor that is a function of the sum of the external momentum $q$ and the internal momentum $k$. We can expand the external momenta assuming that they are small compared to the scale of non-locality as follows
\begin{eqnarray}\label{eq:NL_expansion}
\exp{\Big[ \frac{(k+q)^{2}}{\Lambda^{2}}\Big]} & = & \exp{\Big(\frac{k^{2}+m^{2} + 2 q \cdot k}{\Lambda^{2}}\Big)}, \nonumber\\
& = &  \exp{\Big(\frac{k^{2}+m^{2}}{\Lambda^{2}}\Big)}\Big[ 1+ \frac{2 q\cdot k}{\Lambda^{2}} + \frac{1}{2!} \frac{(2 q\cdot k)^{2}}{\Lambda^{4}} + \frac{1}{3!} \frac{(2 q\cdot k)^{3}}{\Lambda^{6}} + \cdots \Big],\nonumber \\
& = &  \exp{\Big(\frac{k^{2}+m^{2}}{\Lambda^{2}}\Big)}\Big[\cos{\Big( \frac{2i q \cdot k}{\Lambda^{2}}\Big)} - i \sin{\Big( \frac{2i q \cdot k}{\Lambda^{2}}\Big)}  \Big],
\end{eqnarray}
where we have assumed that the fermion is on-shell in the first line. Notice that after the Wick rotation, as required for loop calculations, the exponential factor becomes $\exp{(-k_{E}^{2}/\Lambda^{2})}$, and that the internal momentum is to be integrated over from $0$ to $\infty$, however, in any case, the value of the sine and cosine functions will always be between $0$ and $1$ regardless of the value of $k$. In the regime $k \ll \Lambda$ (with $q \ll \Lambda$ always implied), we can approximate the sine and cosine functions to LO, i.e.
 \begin{eqnarray}\label{eq:1st_regime_approx}
 \cos{\Big( \frac{2 i q \cdot k}{\Lambda^{2}}\Big)} \simeq 1 + O \Big(\frac{1}{\Lambda^{2}} \Big), \nonumber \\
 \sin{\Big( \frac{2 i q \cdot k}{\Lambda^{2}}\Big)} \simeq 0 + O \Big(\frac{1}{\Lambda^{2}} \Big),
 \end{eqnarray}

On the other hand, in the regime where $k \gg \Lambda$ (again with $q \ll \Lambda$ also implied), the exponential factor $e^{-\frac{k^{2}}{\Lambda^{2}}} \ll 1$, which implies that in eq. (\ref{eq:NL_expansion})
 \begin{eqnarray}\label{eq:2nd_regime_approx}
 \cos{\Big( \frac{2 i q \cdot k}{\Lambda^{2}}\Big)} e^{-\frac{k_{E}^{2}}{\Lambda^{2}}},\hspace{2mm} \sin{\Big( \frac{2 i q \cdot k}{\Lambda^{2}}\Big)} e^{-\frac{k_{E}^{2}}{\Lambda^{2}}} \ll 1,
 \end{eqnarray}
and the contribution in this regime is negligible. This only leaves the regime where $k \sim \Lambda$. In this regime, $e^{-\frac{k_{E}^{2}}{\Lambda^{2}}}$ is not small, however, $q \cdot k/\Lambda^{2} \sim q/\Lambda \ll 1$, which implies that the approximation in eq. (\ref{eq:1st_regime_approx}) continues to hold (as long as the $q \ll \Lambda$). Therefore, the non-locality form factor can be approximated as follows
\begin{equation}\label{eq:Final_approx}
\exp{\Big[ \frac{(k+q)^{2}}{\Lambda^{2}}\Big]} \simeq \exp{\Big(\frac{k^{2}+m^{2}}{\Lambda^{2}}\Big)}\Big[ 1 + \frac{2q\cdot k}{\Lambda^{2}} + \cdots \Big].
\end{equation} 

Finally, we point out that given the experimental bound on $\Lambda \gtrsim 2.5 - 3$ TeV, for all practical purposes, the mass in eq. (\ref{eq:Final_approx}) can also be dropped. We have checked numerically that for masses $\sim O(\text{EW})$, the LO in the approximation in eq. (\ref{eq:Final_approx}) differs from the numerical solution be no more than $1\%$ \cite{Abu-Ajamieh:2023syy}, thus justifying our treatment.

\subsection{The Matrix Element at LO in $1/\Lambda$}\label{sec:III_2}
Given the above expansion, we can simplify the matrix element in eq. (\ref{eq:NLO_matrix}). Working to the LO in $1/\Lambda$, it is easy to see that all non-locality factors become $\sim e^{k^{2}/\Lambda^{2}}(1+O(1/\Lambda^{2}))$. Thus, the full matrix element simplifies to 
\begin{eqnarray}\label{eq:expanded_matrix}
i\mathcal{M}^{\mu} & \simeq &  -\frac{e^{3}}{4} \int \frac{d^{4}k}{(2\pi)^{4}}\Bigg[ \frac{\exp{\Big( \frac{3k^{2}}{\Lambda_{f}^{2}}+\frac{k^{2}}{\Lambda_{g}^{2}}\Big)}}{[(k+p)^{2}-m^{2}][(k-q_{1})^{2}-m_{\gamma}^{2}][k^{2}-m^{2}]}\Bigg] \nonumber\\
&\times & \overline{u}(q_{2}) \Bigg\{ \Big(e^{\frac{k^{2}}{\Lambda_{f}^{2}}} + 1 \Big)^{2} \gamma^{\nu}(\slashed{k} + \slashed{p} + m) \gamma^{\mu} (\slashed{k}+m) \gamma_{\nu} \nonumber \\
& - & \Bigg( \frac{e^{\frac{2k^{2}}{\Lambda_{f}^{2}}}-1}{k^{2}-m^{2}} \Bigg)\gamma^{\nu}(\slashed{k} + \slashed{p} + m) \gamma^{\mu} (\slashed{k}+m)(k_{\nu}\slashed{q}_{1}+q_{1\nu}\slashed{k}) \\
& - & \Bigg( \frac{e^{\frac{2k^{2}}{\Lambda_{f}^{2}}}-1}{(k+p)^{2}-m^{2}} \Bigg)[(k^{\nu}+p^{\nu})\slashed{q}_{2}+q_{2}^{\nu}(\slashed{k}+\slashed{p})](\slashed{k} + \slashed{p} + m) \gamma^{\mu}  (\slashed{k}+m)\gamma_{\nu} \nonumber \\
& + & \Big( e^{\frac{k^{2}}{\Lambda_{f}^{2}}}-1\Big)^{2}\frac{[(k^{\nu}+p^{\nu})\slashed{q}_{2}+q_{2}^{\nu}(\slashed{k}+\slashed{p})]}{[(k+p)^{2}-m^{2}][k^{2}-m^{2}]}(\slashed{k} + \slashed{p} + m)\gamma^{\mu} (\slashed{k}+m)(k_{\nu}\slashed{q}_{1}+q_{1\nu}\slashed{k})
\Bigg\}u(q_{1}), \nonumber
\end{eqnarray}
where we have given the photon a mass $m_{\gamma}$ in order to regulate the IR divergence that corresponds to the massless photon. The expression is quite complex. However, it can be greatly simplified by inspecting the naive degree of divergence for each term. It is not hard to see that the first term dominates, whereas the remaining 3 yield subleading corrections. More specifically, we see that the first term $\sim k^{0}$, whereas the second and third terms $\sim 1/k$, and the last $\sim 1/k^{2}$. Thus, the last 3 terms will be suppressed by one or two extra powers of the scale non-locality compared to the first. Explicit calculation shows that the first term yields a contribution $\sim O(1)$, whereas the remaining terms yield a leading contribution $\sim O(m^{2}/\Lambda_{f}^{2})$ at best. Therefore they can be safely neglected. In the following, we only consider the contribution from the first term.

In order to extract the NLO corrections to the form factors $\delta F_{1}(p^{2})$ and $\delta F_{2}(p^{2})$, we need to bring the matrix element to the canonical form
\begin{equation}\label{eq:canonical_form}
i\mathcal{M}^{\mu} = (-i e) \overline{u}(q_{2}) \Bigg[\delta F_{1}\Big( \frac{p^{2}}{m^{2}}\Big)\gamma^{\mu} + \frac{i \sigma^{\mu\nu}}{2m}p_{\nu} \delta F_{2}\Big( \frac{p^{2}}{m^{2}}\Big) \Bigg] u(q_{1}).
\end{equation}

Following the standard procedure of collecting the denominators through the Feynman parameters, then using the Ward and Gordon identities, and after significant Dirac algebra, the matrix element becomes
\begin{eqnarray}\label{eq:canonical_matrix}
i\mathcal{M}^{\mu} & \simeq & e^{3} \int dx dy dz \delta(1-x-y-z) \int \frac{d^{4}k}{(2\pi)^{4}} \Bigg( e^{ (5+r^{2})\frac{k^{2}}{\Lambda_{f}^{2}}} + 2e^{(4+r^{2})\frac{k^{2}}{\Lambda_{f}^{2}}} + e^{(3+r^{2})\frac{k^{2}}{\Lambda_{f}^{2}}} \Bigg) \nonumber\\
& \times & \overline{u}(q_{2}) \Bigg( \frac{-\frac{1}{2}k^{2}+(1-4z+z^{2})m^{2}}{(k^{2}-\Delta)^{3}}\gamma^{\mu}+ \frac{i m z(1-z)}{(k^{2}-\Delta)^{3}} p_{\nu}\sigma^{\mu\nu}\Bigg) u(q_{1}),
\end{eqnarray}
where we have defined $r \equiv \Lambda_{f}/\Lambda_{g}$ and $\Delta = -x y p^{2} +(1-z)^{2} m^{2} + z m_{\gamma}^{2}$. 

\subsection{Extracting $\delta F_{1}(p^{2})$}\label{sec:III_3}
Comparing eq. (\ref{eq:canonical_form}) with eq. (\ref{eq:canonical_matrix}), we find that
\begin{eqnarray}\label{eq:F1_mat}
\delta F_{1}(p^{2}=0) & \simeq & i e^{2}\int dx dy dz \delta(1-x-y-z) \int \frac{d^{4}k}{(2\pi)^{4}} \Big( e^{ (5+r^{2})\frac{k^{2}}{\Lambda_{f}^{2}}} + 2e^{(4+r^{2})\frac{k^{2}}{\Lambda_{f}^{2}}} + e^{(3+r^{2})\frac{k^{2}}{\Lambda_{f}^{2}}} \Big) \nonumber \\
& \times & \Bigg( \frac{-\frac{1}{2}k^{2}+(1-4z+z^{2})m^{2}}{(k^{2}-\Delta)^{3}} \Bigg),
\end{eqnarray}
and it is quite straightforward to calculate the momentum and Feynman integrals to obtain
\begin{eqnarray}\label{eq:F1_unreg}
\delta F_{1} \simeq \frac{\alpha}{16\pi}\Bigg[ 18 - 8\log{\Big( \frac{m^{2}}{m_{\gamma}^{2}}\Big)} - \text{Ei}\Big(\frac{-(3+r^{2})m^{2}}{\Lambda_{f}^{2}} \Big) - 2\text{Ei}\Big(\frac{-(4+r^{2})m^{2}}{\Lambda_{f}^{2}} \Big) - \text{Ei}\Big(\frac{-(5+r^{2})m^{2}}{\Lambda_{f}^{2}} \Big)
\Bigg]
\end{eqnarray}
where the exponential integral function $\text{Ei}(t)$ is given by
\begin{equation}\label{eq:Ei}
\text{Ei}(t) = - \int_{-t}^{\infty}dz \frac{e^{-z}}{z},
\end{equation}
and we have dropped the photon mass inside $\text{Ei}(t)$ since it is free from any IR divergences. Nonetheless, there is a logarithmic IR divergence when $m_{\gamma} \rightarrow 0$ which needs to be regularized, just like the case in local QED. On the other hand, unlike local QED, the result is free from any UV divergences, which is a direct result of the exponential non-locality form factor that rendered all amplitudes finite. Curing the IR divergence can be done by following the prescription used in local QED. For instance, since $m_{\gamma}^{2} = p^{2} = E^{2}$, we can set $E$ to be equal to some minimum energy $E_{\text{min}}$ that defines the resolution of the detector, however, we find it more suitable to use the renormalization prescription formulated in \cite{Abu-Ajamieh:2023syy} to renormalize non-local QFTs.

The reason behind this choice is as follows: Although after removing the IR divergence in eq. (\ref{eq:F1_unreg}) and using $\Lambda_{r} \simeq 3$ TeV that corresponds to solving the $(g-2)_{\mu}$ anomaly, one obtains (assuming $r \sim 1$) an NLO correction $\delta F_{1} \sim O(10^{-2})$, which is acceptable; we should note that this does not hold when the scale of non-locality is much larger than that. Specifically, as found in \cite{Abu-Ajamieh:2023syy}, quadratic divergences in the local QFT translate into a quadratic dependence on the scale of non-locality in the non-local version, and logarithmic divergences in the local QFT translate into a logarithmic-like dependence on the scale of non-locality through the function $\text{Ei}(-t)$, which behaves like a logarithmic function for $t \ll 1$. These threshold correction could be unacceptably large in some situations (as we shall show below), and become divergent when $\Lambda_{f} \rightarrow \infty$.

To remedy such situations, \cite{Abu-Ajamieh:2023syy} formulated a prescription to renormalize non-local QFTs dubbed the Non-locality Renormalization Scheme (NRS), where any quadratic dependence on the scale on non-locality is subtracted, whereas the logarithmic-like dependence is either kept (the Minimum Non-locality Subtraction or $\text{MNS}$), or subtracted (the Modified Minimum Non-locality Subtraction or $\overline{\text{MNS}}$), such that in the limit $\Lambda \rightarrow \infty$, the non-local result will correspond to the dimensionally regularized local case, up to possibly scheme-dependent constant terms. Both schemes are generic and can be used for any QFT. In our case, the quadratic dependence on $\Lambda_{f}$ is already absent, thus we need to utilize the $\overline{\text{MNS}}$ scheme. In the $\overline{\text{MNS}}$ scheme, we use the following prescription
\begin{equation}\label{eq:MNS_bar}
	\text{Ei}\Big(\frac{-m^{2}}{\Lambda^{2}}\Big) \rightarrow \text{Ei}\Big(\frac{-m^{2}}{\Lambda^{2}}\Big) - \text{Ei}\Big(\frac{-\mu^{2}}{\Lambda^{2}}\Big),
\end{equation}
where $\mu$ is a renormalization scale. Using this prescription with eq. (\ref{eq:F1_unreg}), then using the expansion of $\text{Ei}(-t)$ for small argument
\begin{equation}\label{eq:Ei_expansion}
\text{Ei}\Big(\frac{-m^{2}}{\Lambda^{2}}\Big) \simeq \gamma_{E} + \log{\Big(\frac{m^{2}}{\Lambda^{2}}\Big)},
\end{equation}
eq. (\ref{eq:F1_unreg}) simplifies to
\begin{equation}\label{eq:ren_condition}
	\delta F_{1}(0) \simeq \frac{\alpha}{4\pi} \Big[\frac{9}{2} - 2\log{\Big( \frac{m^{2}}{m_{\gamma}^{2}}\Big)} - \log{\Big( \frac{m^{2}}{\mu^{2}}\Big)} \Big].
\end{equation}

The renormalization scale $\mu$, should be chosen appropriately based on some renormalization condition. As suggested in \cite{Abu-Ajamieh:2023syy}, the NRS was introduced so that in the limit $\Lambda \rightarrow \infty$, the renormalized non-local QFT would agree with the renormalized local QFT, upto possible scheme-dependent constant terms. Therefore, a suitable renormaliztion condition is to require the vanishing of all higher-order corrections to $F_{1}$. Therefore, setting eq. (\ref{eq:ren_condition}) to vanish implies
\begin{equation}\label{eq:mu}
\mu^{2} \equiv \frac{m^{6}}{m_{\gamma}^{4}}e^{-9/2},
\end{equation}
and it is fairly easy to see that through suitable renormalization conditions, all higher-orders corrections to $F_{1}$ can be made to vanish, such that the renormalization at tree-level is exact to all orders in non-local QED, just as the case in local QED, i.e.
\begin{equation}\label{eq:F1_total}
F_{1}(0) = F^{\text{LO}}_{1}(0) +  \delta F^{\text{NLO}}_{1}(0) + \delta F^{\text{NNLO}}_{1}(0) + \cdots = F^{\text{LO}}_{1}(0) = \Big(1+\frac{m^{2}}{\Lambda_{f}^{2}}\Big) e^{\frac{m^{2}}{\Lambda_{f}^{2}}} \simeq 1+ \frac{2m^{2}}{\Lambda_{f}^{2}}, 
\end{equation}
where $F^{\text{LO}}_{1}(0)$ is obtained from eq. (\ref{eq:F1_tree}) by taking the limit $q_{2}^{2} \rightarrow q_{2}^{2} = m^{2}$, and we see that $\lim_{\Lambda_{f} \rightarrow \infty} F_{1}(0) = 1$, in agreement with the local limit.

\subsection{Extracting $\delta F_{2}(q^{2})$}\label{sec:III_4}

Comparing eq. (\ref{eq:canonical_matrix}) to eq. (\ref{eq:canonical_form}), we immediately identify the NLO correction to $F_{2}(p^{2})$
\begin{eqnarray}\label{eq:F2_mat}
\delta F_{2}(p^{2}=0) & \simeq & 2i m^{2} e^{2} \int dx dy dz \delta(1-x-y-z)z(1-z)  \nonumber \\
& \times &  \int \frac{d^{4}k}{(2\pi)^{4}} \Bigg( e^{ (5+r^{2})\frac{k^{2}}{\Lambda_{f}^{2}}} + 2e^{(4+r^{2})\frac{k^{2}}{\Lambda_{f}^{3}}} + e^{(3+r^{2})\frac{k^{2}}{\Lambda_{f}^{2}}} \Bigg)  \Bigg( \frac{1}{(k^{2}-\Delta)^{3}} \Bigg),
\end{eqnarray}
and following the same procedure used to obtain $\delta  F_{1}$, it is fairly straightforward to evaluate the integrals and find
\begin{eqnarray}\label{eq:F2}
\delta F_{2}(p^{2}=0) & \simeq &  \frac{\alpha}{2\pi}\Bigg[1 + \frac{(3+r^{2})m^{2}}{12\Lambda_{f}^{2}}\text{Ei}\Big(\frac{-(3+r^{2})m^{2}}{\Lambda_{f}^{2}}\Big)+ \frac{(4+r^{2})m^{2}}{6\Lambda_{f}^{2}}\text{Ei}\Big(\frac{-(4+r^{2})m^{2}}{\Lambda_{f}^{2}}\Big) \nonumber \\
& + & \frac{(5+r^{2})m^{2}}{12\Lambda_{f}^{2}}\text{Ei}\Big(\frac{-(5+r^{2})m^{2}}{\Lambda_{f}^{2}}\Big) + O\Big(\frac{m^{2}}{\Lambda_{f}^{2}}\Big) \Bigg].
\end{eqnarray}

Given the vanishing of $F_{2}(0)$ at tree-level, we can see that eq. (\ref{eq:F2}) yields the correct local limit, i.e. $\lim_{\Lambda_{f} \rightarrow \infty} F_{2}(0) = \alpha/2\pi$. Nevertheless, the result in eq. (\ref{eq:F2}) is inaccurate. More specifically, although the leading term in eq. (\ref{eq:F2}) is indeed correct, the subleading corrections are incomplete. The reason behind this is that the leading corrections arising from the neglected terms in eq. (\ref{eq:expanded_matrix}) (the last 3 terms), and the NLO expansion terms in external momenta $p\cdot k/\Lambda_{f}^{2}$ and $q\cdot k/\Lambda_{f}^{2}$ in the first term in eq. (\ref{eq:expanded_matrix}), will yield contributions of the same size as the subleading corrections in eq. (\ref{eq:F2}). Therefore, a proper calculation of $\delta F_{2}(0)$ would require expanding eq. (\ref{eq:expanded_matrix}) to NLO in the external momenta, then keeping the terms that yield contributions of the same order as the subleading corrections in eq. (\ref{eq:F2}). However, this is a challenging task, as there are literally hundreds of terms in the expansion. Nonetheless, there is no need for alarm, as the size of these contribution is easy to estimate
\begin{equation}\label{F2_NLO_estimate}
\delta F_{2}^{\text{NLO}}(p^{2}=0) \sim k \times \frac{\alpha m^{2}}{16\pi\Lambda_{f}^{2}}\text{Ei}\Big( \frac{-m^{2}}{\Lambda_{f}^{2}}\Big),
\end{equation}
where $k \sim$ (a few). Thus we have
\begin{equation}\label{eq:F2_total}
F_{2}(p^{2}=0) \simeq \frac{\alpha}{2\pi}\Bigg[ 1 + k \frac{m^{2}}{8\Lambda_{f}^{2}} \text{Ei} \Big(\frac{-m^{2}}{\Lambda_{f}^{2}}\Big)\Bigg].
\end{equation}

Numerically, if we use $\Lambda_{f} \sim 3$ TeV as required by the anomaly, and set $k = 100$, then the NLO correction would be $\sim \frac{\alpha}{2\pi} \times O(10^{-7}) \ll F_{2}^{\text{LO}}(0)$. Therefore, the NLO corrections can be safely neglected and there is no need for any exact calculation at NLO.

\section{Further Results at 1-loop in Non-local QED}\label{sec:IV}
We have shown in Section \ref{sec:II} how to correct the results in \cite{Capolupo:2022awe} and provided  the proper treatment to calculate loop diagrams in non-local QFTs in the limit $\Lambda \gg p_{\text{ext}}$. In this section, we will utilize this treatment in order to perform several novel calculations in non-local QED. Specifically, we calculate the photon self-energy, how non-locality modifies the Coulomb potential of electric charges, and the modification of non-locality to the enegy levels of the hydrogen atom, especially its contribution to the Lamb shift. We also discuss charge dequantization through non-locality and the corresponding bounds on the scale of non-locality that can be obtained from millicharge searches.

\subsection{Photon Self-energy}\label{sec:IV_1}
\begin{figure}[!t]
\centerline{\begin{minipage}{0.75\textwidth}
\centerline{\includegraphics[width=300pt]{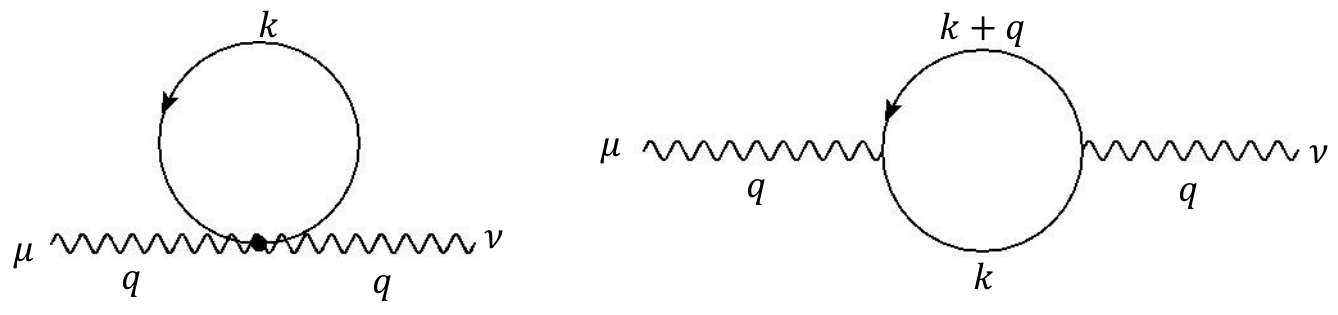}}
\caption{\small Photon self-energy in non-local QED. The diagram on the left is unique to the non-local case and arises from expanding the covariant derivative in the non-local form factor.}
\label{fig2}
\end{minipage}}
\end{figure}
Here we calculate the photon self-energy in non-local QED. This result is necessary for calculating the corresponding modification to the Coulomb potential, i.e. the non-local Uehling potential. Unlike local QED, where there is only one diagram that contributes to the photon self-energy at 1-loop (the diagram on the RHS of Figure \ref{fig2}), non-local QED has an additional contribution through the vertex $\overline{\psi}\psi\gamma\gamma$ (the diagram on the LHS of Figure \ref{fig2}). This diagram arises from expanding the covariant derivative in the non-local form factor and is a direct result of requiring the action to be gauge invariant. The contribution from this diagram is given by
\begin{equation}\label{eq:SE_1}
i\Pi^{\mu\nu} = -i e^{2}\int \frac{d^{4}k}{(2\pi)^{4}}\text{Tr}\Big[ V^{\mu\nu} \frac{ie^{\frac{k^{2}}{\Lambda_{f}^{2}}}(\slashed{k}+m)}{k^{2}-m^{2}} \Big],
\end{equation} 
where $V^{\mu\nu}$ corresponds to the vertex $\overline{\psi}\psi\gamma\gamma$. The Feynman rule that corresponds to this vertex is a very complicated function that was calculated in \cite{Abu-Ajamieh:2023roj}. Luckily, we don't need to evaluate this contribution explicitly, as it can easily be shown that when closing the loop (i.e. when the momenta of the fermions are equal), $V^{\mu\nu}$ vanishes. On the other hand, the second contribution is non-vanishing and to the leading order in $q/\Lambda_{f}$, is given by
\begin{equation}\label{eq:SE_2}
i\Pi^{\mu\nu} = - e^{2}\int \frac{d^{4}k}{(2\pi)^{4}} e^{\frac{4k^{2}}{\Lambda_{f}^{2}}}\text{Tr}\Bigg[ \frac{\gamma^{\mu}(\slashed{k}+\slashed{q}+m)\gamma^{\nu}(\slashed{k}+m)}{((k+q)^{2}-m^{2})(k^{2}-m^{2})} \Bigg],
\end{equation}
which can be calculated through the standard way, yielding the result
\begin{equation}\label{eq:SE_3}
\Pi^{\mu\nu} \simeq - \frac{2\alpha}{\pi} \int_{0}^{1} dx \Bigg[ \frac{\Lambda_{f}^{2}}{16} g^{\mu\nu} +(q^{2}g^{\mu\nu}-q^{\mu}q^{\nu})x(x-1)\text{Ei}\Big( \frac{-4m^{2}-4x(x-1)q^{2}}{\Lambda_{f}^{2}}\Big)
\Bigg] + O(1).
\end{equation}

Notice that the result has a quadratic dependence on the $\Lambda_{f}^{2}$, as well as a logarithmic-like  dependence on the $\Lambda_{f}^{2}$ through the function $\text{Ei}(t)$. As explained in detail in \cite{Abu-Ajamieh:2023syy} and highlighted in section \ref{sec:III_3}, this dependence on the non-local scale, corresponds to the quadratic and logarithmic divergences in the local theory (when a UV cutoff is used). Although the quadratic and logarithmic dependence imply that the non-local version of QED is finite, they nonetheless introduce unacceptably large threshold corrections as we show later on, and thus they need regularization. Following the NRS prescription introduced in \cite{Abu-Ajamieh:2023syy}, any quadratic dependence on the scale of non-locality must be subtracted, whereas the logarithmic-like dependence may or may not be subtracted depending on the scheme, (see discussion around eq. (\ref{eq:MNS_bar})). The quadratic dependence on  $\Lambda_{f}$ can be straightforwardly subtracted. On the other hand, to subtract the logarithmic-like dependence on the scale on non-locality, we modify the $\overline{\text{MNS}}$ slightly to make closer to the local case, i.e. we subtract the divergence as follows
\begin{equation}\label{eq:Log_subtraction}
\widehat{\Pi}_{2}(q^{2}) = \Pi_{2}(q^{2}) - \Pi_{2}(0),
\end{equation}
which can always be arranged via a suitable choice of $\mu^{2}$ in eq. (\ref{eq:MNS_bar}). Thus, the renormalized photon self-energy reads
\begin{eqnarray}
\widehat{\Pi}^{\mu \nu}(q^{2}) & \simeq & (q^{2}g^{\mu\nu}-q^{\mu}q^{\nu}) \widehat{\Pi}_{2}(q^{2}),\\
& \text{with,} & \nonumber \label{eq:reg_SE1} \\
\widehat{\Pi}_{2}(q^{2}) & \simeq & \begin{cases}
-\frac{2\alpha}{\pi} \int_{0}^{1}dx x(x-1)\text{Ei}\Big( \frac{-4m^{2}-4x(x-1)q^{2}}{\Lambda_{f}^{2}}\Big), \hspace{32mm} \text{MNS};  \label{eq:reg_SE2}\\
-\frac{2\alpha}{\pi} \int_{0}^{1}dx x(x-1)\Big[ \text{Ei}\Big( \frac{-4m^{2}-4x(x-1)q^{2}}{\Lambda_{f}^{2}}\Big) - \text{Ei}\Big( \frac{-4m^{2}}{\Lambda_{f}^{2}}\Big)  \Big], \hspace{5mm} \overline{\text{MNS}}.
\end{cases}
\end{eqnarray}

As we shall see below, it is necessary to renormalize the photon self-energy this way, otherwise, large threshold corrections will render the theory essentially nonphysical. Notice from eq. (\ref{eq:reg_SE1}) that photon self-energy in non-local QED preserves both the Lorentz invariance and the Ward identity, and that the photon remains massless, just like the local case. This is reasonable as the action in eq. (\ref{eq:NL_QED_Lag}) is both Lorentz and gauge invariant by construction.

\subsection{Modification to the Electric Potential}\label{sec:IV_2}

It is well-known in local QED, that $\widehat{\Pi}_{2}(q^{2})$ will modify the electric potential between two electric point charges compared to the classical Coulomb potential. In general, the modified electric potential can be expressed as
\begin{equation}\label{eq:V_modification}
V(\vec{r}) = -e^{2}\int \frac{d^{3}\vec{q}}{(2\pi)^{3}} \frac{ e^{i\vec{q} \cdot \vec{r}}}{|\vec{q}|^{2}\Big( 1-\widehat{\Pi}_{2}(-|\vec{q}|^{2})\Big)},  
\end{equation}
and in local QED, this modifies the the electric potential by the Uehling term\footnote{This Uehling potential is an approximation. We show a more careful treatment in Appendix \ref{sec:appendix}.}
\begin{equation}\label{eq:Uehling1}
V(\vec{r}) = -\frac{\alpha}{|\vec{r}|} -\frac{4\alpha^{2}}{15m^{2}}\delta^{3}(\vec{r}).
\end{equation}

Here, we seek to calculate the modified electric potential in non-local QED. Thus, we use $\widehat{\Pi}_{2}(q^{2})$ given in eq. (\ref{eq:reg_SE2}) to calculate the modified potential in eq. (\ref{eq:V_modification}). We first use the $\text{MNS}$ result in our calculation. In the non-relativistic limit momentum $|\vec{q}|^{2} \ll m^{2}$, and with $m \ll \Lambda_{f}$, we have
\begin{equation}\label{eq:Ei_small_q}
\text{Ei}\Big( \frac{-4m^{2}+4x(x-1)|\vec{q}|^{2}}{\Lambda_{f}^{2}}\Big) \simeq \text{Ei}\Big(\frac{-4m^{2}}{\Lambda_{f}^{2}}\Big)  - x(x-1) \frac{|\vec{q}|^{2}}{m^{2}},
\end{equation}
and the integral in eq. (\ref{eq:reg_SE2}) is easily evaluated, yielding
\begin{equation}
\widehat{\Pi}_{2}^{\text{MNS}}(-|\vec{q}|^{2}) \simeq \frac{\alpha}{3\pi}\Big[\text{Ei}\Big(\frac{-4m^{2}}{\Lambda_{f}^{2}}\Big) + \frac{|\vec{q}|^{2}}{5m^{2}} \Big],
\end{equation}
which can be used in eq. (\ref{eq:V_modification}) to find the modified electric potential
\begin{equation}\label{eq:V_modification_NL_MNS}
V_{\text{MNS}}(\vec{r}) \simeq \frac{1}{1-\frac{\alpha}{3\pi}\text{Ei}\Big(\frac{-4m^{2}}{\Lambda_{f}^{2}}\Big)} \Bigg[-\frac{\alpha}{|\vec{r}|}-\frac{4\alpha^{2}}{15m^{2}} \frac{1}{1-\frac{\alpha}{3\pi}\text{Ei}\Big(\frac{-4m^{2}}{\Lambda_{f}^{2}}\Big)} \delta^{3}(\vec{r})\Bigg].
\end{equation}

\begin{figure}[!t]
\centerline{\begin{minipage}{0.75\textwidth}
\centerline{\includegraphics[width=300pt]{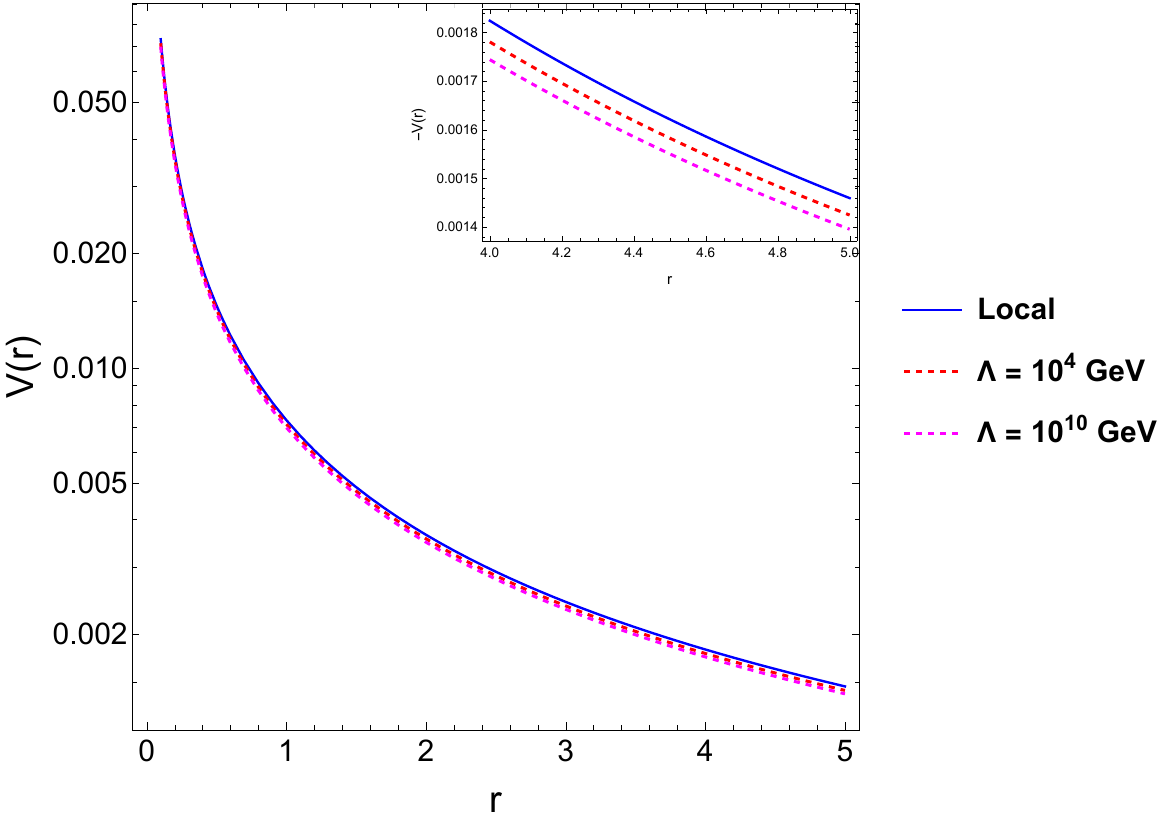}}
\caption{\small Comparison between the leading terms in the modified potential in local and non-local QED in the MNS scheme for 2 benchmark points for the scale of (fermionic) non-locality; $\Lambda_{f} =  10^{4}$ GeV and $\Lambda_{f} =  10^{10}$ GeV.}
\label{fig3}
\end{minipage}}
\end{figure}

We compare between the local and the non-local Uehling potentials for two benchmarks in Figure \ref{fig3}. Notice that non-locality always \textit{lowers} the potential, where we see that the non-local potential is $\sim 2.4\%$ ($\sim 4.4\%$) lower than the local case for $\Lambda_{f} =  10^{4}$ ($10^{10}$) GeV. The reason for this is that non-locality suppressed contributions from higher momenta, thereby providing additional screening to the electric charge. 

Although the modified non-local Coulomb potential exhibits similar characteristics to the local one as expected, it nonetheless suffers from a serious anomaly: For any EFT to be physically sound, the UV sector should yield IR corrections that are represented by irrelevant operators suppressed by the scale of the UV sector, which become smaller the larger the UV scale becomes, eventually decoupling completely when the UV scale is sent to infinity. In other words, for our non-locality description to constitute a physically sound EFT, the corrections of the non-locality in the IR should becomes \textit{smaller} as $\Lambda_{f}$ becomes larger, eventually restoring the local limit when $\Lambda_{f} \rightarrow \infty$. However, what we are witnessing here is quite the opposite behavior: The IR effects of non-locality becomes \textit{larger} as $\Lambda_{f}$ increases (i.e. non-locality reduces the Coulomb potential more as its scale increases), and when $\Lambda_{f} \rightarrow \infty$, the electric potential vanishes instead of reproducing the local limit! 

The reason behind this anomalous behavior lies in the logarithmic-like dependence on the scale of non-locality, which is encoded in the function $\text{Ei}(-m^{2}/\Lambda_{f}^{2})$. As explained above, the logarithmic divergence in the local case, which is removed via the essentially ad hoc prescription in eq. (\ref{eq:Log_subtraction}), is retained in the non-local $\text{MNS}$ scheme because non-locality makes it finite, however this function can lead to sizable threshold corrections when $m/\Lambda_{f} \ll 1$ that eventually turn into the (local) logarithmic divergence when $m/\Lambda_{f} \rightarrow 0$. 

Things are quite different when we use $\widehat{\Pi}_{2}(q^{2})$ in the $\overline{\text{MNS}}$ scheme instead (bottom line in eq. (\ref{eq:reg_SE2})) to calculate the modified potential. In this case we find
\begin{equation}\label{eq:V_modification_NL_MNS_bar}
V_{\overline{\text{MNS}}}(\vec{r}) \simeq -\frac{\alpha}{|\vec{r}|} - \frac{4\alpha^{2}}{15m^{2}}e^{-\frac{4m^{2}}{\Lambda_{f}^{2}}}\delta^{3}(\vec{r}),
\end{equation}
and we can see now that this modified potential looks more reasonable as an EFT. We also notice that taking the limit $\Lambda_{f} \rightarrow \infty$ yields the local case in eq. (\ref{eq:Uehling1}). This potential only modifies the local limit marginally, as $\Lambda_{f} \gg m$ from experimental measurements.

\subsection{Corrections to the Hydrogen Energy Levels and Contribution to the Lamb Shift}\label{sec:IV_3}
Corrections to the electric potential will affect the energy levels of the hydrogen atom. Here, we calculate the modified energy levels of the hydrogen atom due to the modified potential found in Section \ref{sec:IV_2}. In particular, we calculate the contribution to the Lamb shift, which can be compared to the experimentally measured value to set limits on the scale of non-locality. The shift in the energy levels that corresponds to the modified potential is given by\footnote{Notice here that $\psi$ is the unperturbed wavefunction corresponding to the local potential. In general, non-locality will induce a small correction $\delta \psi$ to this wavefunction, and the Lamb shift will receive a correction $\sim \delta \psi \delta V \psi$, however, as $\delta \psi$ is the same order as $\delta V$, this correction is second order compared to the leading correction to Lamb shift and for energies $\ll \Lambda_{f}$ relevant for the Lamb shift, it is negligible.}

\begin{equation}\label{eq:energy_levels_shift1}
\Delta E = \int d^{3}\vec{r} \psi^{*}(\vec{r}) \delta V(\vec{r}) \psi(\vec{r}).
\end{equation} 

First, let's use the modified non-local potential in the $\text{MNS}$ scheme, which is given in eq. (\ref{eq:V_modification_NL_MNS}). This potential leads to the following shift in the energy levels
\begin{equation}\label{eq:energy_levels_shift2}
\Delta E = \int d^{3}\vec{r} |\psi(\vec{r})|^{2} \Bigg[-\frac{\alpha}{|\vec{r}|} \frac{1}{1-\frac{\alpha}{3\pi}\text{Ei}\Big(\frac{-4m^{2}}{\Lambda_{f}^{2}}\Big)}-\frac{4\alpha^{2}}{15m^{2}} \frac{1}{\Big[1-\frac{\alpha}{3\pi}\text{Ei}\Big(\frac{-4m^{2}}{\Lambda_{f}^{2}}\Big)\Big]^{2}} \delta^{3}(\vec{r})\Bigg].
\end{equation} 
Notice that the first term is equal to the local case multiplied by a non-local correction factor, thus we can be immediately read it off from the local case
\begin{equation}\label{eq:detla_E1}
\Delta E_{(1)} \simeq \frac{-13.6}{n^{2}\Big[ 1-\frac{\alpha}{3\pi}\text{Ei}\Big(\frac{-4m^{2}}{\Lambda_{f}^{2}}\Big)\Big]} \hspace{2mm}  \text{eV}.
\end{equation}
This term does not contribute to the Lamb shift, which measures the correction to the transition $2S_{1/2} - 2P_{1/2}$. However, it does modify the energy levels themselves as we shall discuss below. The second term is easily evaluated on the support of the delta function 
\begin{equation}\label{eq:detla_E2}
\Delta E_{(2)} \simeq - \frac{4\alpha^{2}}{15m^{2}} \frac{|\psi(0)|^{2}}{\Big[1-\frac{\alpha}{3\pi}\text{Ei}\Big(\frac{-4m^{2}}{\Lambda_{f}^{2}}\Big)\Big]^{2}}.
\end{equation}

Notice that only $S$-wave states are nonzero at the origin. Therefore for $2S$, the wavefunction is $|\psi(0)_{2S}|^{2} = \alpha^{3}m^{3}/8\pi$, which implies that the non-local correction to the Lamb shift is given by
\begin{equation}\label{eq:NL_Lamb_Shift}
\Delta E_{\text{NL}} = -\frac{\alpha^{5}m}{30\pi}\frac{1}{\Big[1-\frac{\alpha}{3\pi}\text{Ei}\Big(\frac{-4m^{2}}{\Lambda_{f}^{2}}\Big)\Big]^{2}} = \frac{\Delta E_{\text{L}}}{\Big[1-\frac{\alpha}{3\pi}\text{Ei}\Big(\frac{-4m^{2}}{\Lambda_{f}^{2}}\Big)\Big]^{2}},
\end{equation}
where $\Delta E_{L}$ is the Lamb shift in local QED. For the transition $2S_{1/2} - 2P_{1/2}$, the difference between the theoretical and experimental energy differences is given by \cite{Eides:2000xc}
\begin{equation}\label{eq:Lamb_shit_diff}
	\delta E = \Delta E^{\text{Th}} - \Delta E^{\text{Exp}} = 1057.833 - 1057.845 = - 0.012 \hspace{2mm}\text{MHz},
\end{equation}
which can be used in conjunction with eq. (\ref{eq:NL_Lamb_Shift}) to set a lower bound on the scale of non-locality. A $2\sigma$ limit can be obtained by requiring 
\begin{equation}\label{eq:Lamb_bound}
|\Delta E_{\text{NL}}| \hspace{1mm} \lesssim \hspace{1mm} |2\delta E|,
\end{equation}
however, the resulting lower bound is ridiculously large $\sim O(10^{13060})$!, and many orders of magnitude larger than the Landau pole. This would clearly exclude non-local QED (at least as prescribed by the action in eq. (\ref{eq:NL_QED_Lag})) as a realistic QFT. To make matters worse, the modification to the energy levels given by eq. (\ref{eq:detla_E1}) will set a different limit on the scale of non-locality. As the energies of transitions in the hydrogen atom are well-measured, they can be used to set (different) limits on the scale on non-locality. The best measures transition we are aware of is the $1S-2S$ transition, which theoretically corresponds to $10.2$ eV. The experimentally measured value of this transition reads \cite{Matveev:2013orb}
\begin{equation}\label{eqL1s2s_measured}
	f_{1S-2S} = 2466061413187018 \pm 11 \hspace{2mm} \text{Hz} \hspace{5mm} \implies \Delta E_{1S-2S} = 10.1988 \pm O(10^{-13}) \hspace{2mm} \text{eV}.
\end{equation}

Using this limit in eq. (\ref{eq:detla_E1}) requires $\Lambda_{f} = 1$ MeV! Clearly this is inconsistent with the other bound and is too low to make any physical sense, not to mention that it is in conflict with the LHC bounds \cite{Biswas:2014yia}. Just as the case with the modified potential, the root of these conflicting results lies the large threshold corrections arising from function $\text{Ei}(-4m^{2}/\Lambda_{f}^{2})$. As illustrated in detail in \cite{Abu-Ajamieh:2023syy}, and as the case with the modified potential above, in some cases, it is not sufficient just to subtract the quadratic dependence on $\Lambda_{f}$, i.e. use the $\text{MNS}$ scheme, and one has to also subtract the logarithmic-like dependence on $\Lambda_{f}$ through the $\overline{\text{MNS}}$ prescription in eq. (\ref{eq:MNS_bar}). Therefore, we should use $\overline{\text{MNS}}$ result for $\widehat{\Pi}_{2}(q^{2})$ in eq. (\ref{eq:reg_SE2}), which in the non-relativistic limit $q^{2} \ll m^{2}$ yields the result\footnote{Notice that in eq. (\ref{eq:Ei_small_q}), we have droped the exponential factor since $m \ll \Lambda_{f}$. However, here we have to keep it as it is the only source of non-locality modification that remains in the $\overline{\text{MNS}}$ scheme.}
\begin{equation}\label{eq:Pi2_MNS_bar}
\widehat{\Pi}_{2}^{\overline{\text{MNS}}} \simeq \frac{\alpha |\vec{q}|^{2}}{15\pi m^{2}} e^{-\frac{4m^{2}}{\Lambda_{f}^{2}}}.
\end{equation}

Using this in eq. (\ref{eq:V_modification}) to evaluate the modified electric potential, we obtain
\begin{equation}\label{eq:V_MNS_bar}
V_{\overline{\text{MNS}}}(|\vec{r}|) = -\frac{\alpha}{|\Vec{r}|} - \frac{4\alpha^{2}}{15m^{2}}e^{-\frac{4m^{2}}{\Lambda_{f}^{2}}}\delta^{3}(\vec{r}),
\end{equation}
and we can see clearly that the potential is not modified away from the origin, and that the local limit in eq. (\ref{eq:Uehling1}) is retrieved when $\Lambda_{f} \rightarrow \infty$. Using this modified potential in eq. (\ref{eq:energy_levels_shift1}), we can easily see that the energy levels will be identical to the local case, since the leading term in the $\overline{\text{MNS}}$ is identical to the local case, thereby evading any bounds from energy transitions. On the other hand, the contribution to the Lamb shift becomes
\begin{equation}\label{eq:Lamb_MNS_bar}
\Delta E_{\overline{\text{MNS}}} \simeq -\frac{\alpha^{5} m}{30\pi}e^{-\frac{4m^{2}}{\Lambda_{f}^{2}}},
\end{equation}
which in conjunction with eq. (\ref{eq:Lamb_shit_diff}) brings the lower limit on $\Lambda_{f}$ from above the Landau pole to a mere $\sim 50$ MeV. 

The $\overline{\text{MNS}}$ scheme results are consistent and logical. In addition, the modified potential and the corresponding Lamb shift yield the correct (local) limit when $\Lambda_{f} \rightarrow \infty$, unlike the case with the $\text{MNS}$ scheme. This means that the $\overline{\text{MNS}}$ is indeed the correct prescription to use for the Lamb shift. Nonetheless, the vast disparity between the results of the two schemes, and the fact that the non-locality formulation does not completely eliminate the need for renormalization in many cases, both merit further comments. We will comment on the results in Section \ref{sec:IV_4} below.

\subsection{Comments on Non-local QED and Non-locality Renormalization}\label{sec:IV_4}
As discussed in detail in \cite{Abu-Ajamieh:2023syy}, although non-locality does indeed render UV-divergent quantities in local QFTs finite, it nonetheless could lead to unacceptably large threshold corrections. The quadratic divergence in the local case translates into a quadratic dependence on the scale of non-locality, whereas the logarithmic divergence in the local case translates into a logarithmic-like dependence on the scale of non-locality through the function $\text{Ei}(t)$. 

The calculation in \ref{sec:IV_3}, shows how unacceptably large these threshold corrections could be. We saw that when the logarithmic-like dependence is kept, non-locality could yield physically inconsistent results. On the other hand, when this problematic dependence is subtracted via the $\overline{\text{MNS}}$ scheme, the results are consistent and yield the correct local limit. 

These calculations seem to suggest that whenever the local version of the theory has a logarithmic divergence, the $\overline{\text{MNS}}$ scheme should be used to avoid large threshold corrections that could contradict the experimental results. Or perhaps one should choose the appropriate renormalization scheme depending on the calculation at hand in an ad-hoc manner, however, in doing so, one faces the obvious question: What is the point of using non-locality to begin with if the results still need regularization to remove any large threshold corrections that could be unacceptable experimentally? 

Non-locality through infinite derivatives as shown in eq. (\ref{eq:NL_form factor}) was mainly introduced as a solution to the UV divergences in the local QFT, i.e. to make the theory super-renormalizable without the need for any renormalization schemes as in local QFTs. Before the age of the LHC, the scale on non-locality was thought (or hoped) to be low, however, given the LHC bound on $\Lambda$ \cite{Biswas:2014yia}, we now know that this is not the case. Thus, any threshold corrections that correspond to UV divergences in the local case will indeed be sizable and even too large in some cases as we saw with the modification to the energy levels of the hydrogen atom and the Lamb shift.

The necessity to use a renormalization scheme to eliminate the non-local "divergence-like" threshold corrections seems to defeat the whole purpose of using non-local QFTs, which casts serious doubt on their usefulness. We will neither attempt to answer this question here, nor suggest that physicists should give up on non-locality. However, what we hope to do is start a discussion on this issue. We will not discuss this issue any further.

\subsection{Charge Dequantization}\label{sec:IV_5}

An important result highlighted in \cite{Capolupo:2022awe} but not emphasized enough, is the fact that non-locality can lead to charge dequantization. It was found in \cite{Capolupo:2022awe} (see eq. (\ref{eq:F1_tree})) that at tree-level
\begin{equation}\label{eq:F1_repeated}
F_{1}(0) = \Big(1+\frac{m^{2}}{\Lambda_{f}^{2}}\Big)e^{\frac{m^{2}}{\Lambda_{f}^{2}}}.
\end{equation}

As we've argued in \ref{sec:III_3} (see eq. (\ref{eq:F1_total}) above), all higher order corrections to $F_{1}(0)$ can be made to vanish through appropriate renormalization conditions. Therefore, $F_{1}(0)$ above is  exact to all orders. Since $F_{1}(0)$ essentially renormalizes the electric charge, i.e.
\begin{equation}\label{eq:Q_dequantization}
F_{1}(0) = Q = \Big(1+\frac{m_{f}^{2}}{\Lambda_{f}^{2}}\Big)e^{\frac{m_{f}^{2}}{\Lambda_{f}^{2}}} \simeq 1+ \frac{2m_{f}^{2}}{\Lambda_{f}^{2}} + \cdots,
\end{equation}
we can see that non-locality leads to charge dequantization. It has long been known that the electric charge can be dequantized via the introduction of a new gauged $U(1)$ group \cite{Babu:1989tq, Babu:1989ex, Foot:1992ui}, however, non-locality now offers a novel way for charge dequantization. 

Notice from eq. (\ref{eq:Q_dequantization}) that the electric charge depends on the ratio $2m_{f}/\Lambda_{f}$, which was interpreted in \cite{Capolupo:2022awe} as either a violation of gauge invariance, or a violation of lepton charge universality. This interpretation is erroneous, since the action in eq. (\ref{eq:NL_QED_Lag}) is gauge invariant by construction, and in all calculation gauge invariance is never broken. What this result truly means is that eq. (\ref{eq:Q_dequantization}) offers 2 possibilities: either the scale of non-locality is universal, which implies a violation of the electric charge universality; or that the electric charge is universal (by keeping the ratio $2m_{f}/\Lambda_{f}$ fixed), which implies a flavor-dependent scale of non-locality.

Let's discuss these two possibilities in more detail. First, Let's assume for the moment that the lepton's charge is universal. From \cite{Foot:1992ui} (see also \cite{Das:2020egb} for more details), it is shown that if the SM is extended by 3 \textit{Dirac} neutrinos, and if the SM fermions obtain their mass through the Higgs mechanism, then charge can be dequantizaed as follows
\begin{eqnarray}
Q^{i}_{\nu_{L}} & = &  Q^{i}_{\nu_{R}} = \epsilon \hspace{5mm} \forall i, \label{eq:neutrino_charge}\\
Q^{i}_{e_{L}} & = & Q^{i}_{e_{R}} = -1 + \epsilon \hspace{5mm} \forall i, \label{eq:lepton_charge}\\
Q^{i}_{u_{L}} & = & Q^{i}_{u_{R}} = \frac{2}{3} + \frac{1}{3}\epsilon \hspace{5mm} \forall i, \label{eq:up_charge}\\
Q^{i}_{d_{L}} & = & Q^{i}_{d_{R}} = -\frac{1}{3} + \frac{1}{3}\epsilon \hspace{5mm} \forall i, \label{eq:down_charge}
\end{eqnarray}
where $i$ refers to the fermion generation. These conditions are obtained by demanding that the SM fermion Yukawa interactions be gauge invariant, and that the gauge anomalies cancel. Comparing this with eq. (\ref{eq:Q_dequantization}), we can immediately identify\footnote{Notice that the eq. (\ref{eq:Q_dequantization}) implies that the charge of any lepton is $=1+\epsilon$, however, eq. (\ref{eq:neutrino_charge}) implies that $Q_{\nu} = \epsilon$. The resolution for this apparent contradiction is to define $Q_{\nu} \equiv F_{1}(0)-1$, since in the local limit, neutrinos do no couple to the photon. This way the local limit is obtained when $\Lambda_{f}\rightarrow\infty$, and the charge dequantization is correct.}
\begin{equation}\label{eq:epsilon}
\epsilon = \frac{2m_{l}^{2}}{\Lambda_{l}^{2}} = \frac{4 m_{u}^{2}}{\Lambda_{u}^{2}} = \frac{2m_{d}^{2}}{\Lambda_{d}^{2}},
\end{equation}
and we can use the limits on the neutrino charges to set bounds on the flavor-dependent scales on non-locality. The most stringent constrains on the charge of neutrinos arises from beta decay in combination with limits on matter neutrality \cite{Foot:1992ui, Giunti:2014ixa, Raffelt:1999gv}
\begin{equation}\label{eq:epsilon_limit}
	|\epsilon| < 10^{-21},
\end{equation}
which translates into the flavor-dependent lower limits on the scale of non-locality for the SM fermions
\begin{eqnarray}
\Lambda_{e} & \gtrsim & 2.29 \times 10^{4} \hspace{2mm} \text{TeV}, \label{eq:e_bound}\\
\Lambda_{\mu} & \gtrsim & 4.74 \times 10^{6} \hspace{2mm} \text{TeV}, \label{eq:mu_bound}\\
\Lambda_{\tau} & \gtrsim & 7.95 \times 10^{7} \hspace{2mm} \text{TeV}, \label{eq:tau_bound}\\
\Lambda_{u} & \gtrsim & 1.37 \times 10^{5} \hspace{2mm} \text{TeV}, \label{eq:u_bound}\\
\Lambda_{d} & \gtrsim & 2.09 \times 10^{5} \hspace{2mm} \text{TeV}, \label{eq:d_bound}\\
\Lambda_{s} & \gtrsim & 4.16 \times 10^{6} \hspace{2mm} \text{TeV}, \label{eq:s_bound}\\
\Lambda_{c} & \gtrsim & 8.03 \times 10^{7} \hspace{2mm} \text{TeV}, \label{eq:c_bound}\\
\Lambda_{b} & \gtrsim & 1.87 \times 10^{8} \hspace{2mm} \text{TeV}, \label{eq:b_bound}\\
\Lambda_{t} & \gtrsim & 1.09 \times 10^{10} \hspace{2mm} \text{TeV}, \label{eq:t_bound}
\end{eqnarray}
and we see that in this case (where neutrinos are assumed to be Dirac fermions), we can obtain stringent bounds that are much stronger than the LHC bounds obtained in \cite{Biswas:2014yia}. On the other hand, if neutrinos are Majoranan fermions, then the invariance of the Weinberg operator implies that they must be electrically neutral \cite{Babu:1989tq, Babu:1989ex, Foot:1992ui}, which according to the conditions in eq. (\ref{eq:neutrino_charge})-(\ref{eq:down_charge}) implies that the electric charge of the SM is quantized. Therefore, non-local QED with infinite derivatives as formulated in the action (\ref{eq:NL_QED_Lag}) is inconsistent with Majorana neutrinos.

Finally, we discuss the second possibility represented by a universal non-locality scale. We can immediately see that this possibility is excluded. The reason for this is that a universal non-locality scale implies a flavor-dependent charge, which cannot be accommodated according to the conditions in eqs. (\ref{eq:neutrino_charge}) - (\ref{eq:down_charge}). Therefore, we conclude that non-locality cannot be universal and must be flavor-dependent (assuming it exists!).

Before we conclude this section, we should point out that although it appears that the renormalization condition in eq. (\ref{eq:F1_total}) can be used to set the electric charge $Q=1$ and thus avoid charge dequantization altogether; this choice is actually not possible, as it would involve tuning all higher-order corrections against the tree-level one to reproduce the desired value. This tuning would include a complicated dependence on the different scales of non-locality of all of the particles running in the higher-order loops, which come at different powers and should cancel against each other. Satisfying such a condition will not be possible.

\section{Discussion and Conclusions}\label{sec:V}
In this paper, we corrected the results obtained in \cite{Capolupo:2022awe} and provided the correct treatment for extracting the magnetic dipole moment in non-local QED. In specific, we showed how to properly extract the form factors $F_{1}(p^{2})$ and $F_{2}(p^{2})$, and showed how to properly regularize them using the NRS scheme proposed in \cite{Abu-Ajamieh:2023syy}. We also showed how non-local form factors exhibit the correct local limit when the scale of non-locality $\Lambda_{f} \rightarrow \infty$.

We used our treatment to calculate some novel results in non-local QED. Specifically, we calculated the photon self-energy and showed that in non-local QED, the result is both Lorentz and gauge invariant, which is a reasonable result given that the non-local QED action is both Lorentz and gauge invariant by construction. This means that the photon remains massless and the Ward identity is respected in non-local QED, just as the case with local QED. Part of these results was also demonstrated in \cite{Abu-Ajamieh:2023roj}, and they contradict the claim made in \cite{Capolupo:2022awe} that gauge invariance could be broken.

We also calculated the non-local Uehling potential in both the $\text{MNS}$ and $\overline{\text{MNS}}$ schemes. We found that in the former case, the electric potential is indeed modified at and away from the origin, where non-locality always enhances the charge screening effect, thereby lowering the potential compared to the local limit. Nonetheless, we found counter-intuitively, that increasing the scale of non-locality would enhance this effect, in contradiction to how a sound EFT should behave. We showed that the reason behind this lies in the logarithmic-like "divergence" which, although being absent in the local case, gets included in the non-local case. We showed that subtracting the logarithmic-like dependence on the scale of non-locality through the $\overline{\text{MNS}}$ scheme remedies the situation, and we showed that in that case non-locality almost has no impact on the electric potential away from the origin, and only marginally modifies the potential at the origin. 

We also calculated the modification of the energy levels of the hydrogen atom and the contribution to the Lamb shift due to non-locality. We found that when the $\text{MNS}$ scheme is used, experimental constraints on the measured energy levels of the hydrogen atom and the Lamb shift are inconsistent and essentially rule out non-locality as a viable prescription. On the other hand, we found that when the $\overline{\text{MNS}}$ scheme is used, i.e. when the logarithmic-like dependence on the scale of non-locality is subtracted, the situation is cured. Specifically, we found that using the $\overline{\text{MNS}}$ scheme, there will be no modification to the energy levels compared to the local case, whereas the modification to the Lamb shift due to non-locality is marginal and sets a lower limit of only $\sim 50$ MeV.

Throughout our calculation, we showed that although non-locality does indeed render any UV-divergent quantities in the local case finite, it nonetheless could introduce large threshold corrections that could be problematic, and could even contradict experimental bounds. In specific, we showed that quadratic divergences in local QFTs translate into a (finite) quadratic dependence on the scale of non-locality, whereas logarithmic divergences in local QFTs translate into a (finite) logarithmic-like dependence on the scale on non-locality through the function $\text{Ei}(t)$. We argued that this situation can be remedied by utilizing the NRS prescription formulated in \cite{Abu-Ajamieh:2023syy}, whereby the quadratic and logarithmic dependence can be subtracted following a suitable subtraction scheme, in a manner similar to dimensional regularization in local QFTs, however, we also argued that the need for regularizing non-local QFTs brings into question their usefulness as a means to regularize (divergent) local QFTs. We are hoping that our findings would bring a serious discussion vis-a-vis the correct prescription to formulate non-local QFTs, or whether non-locality should be incorporated into QFTs to begin with.

We also showed that non-locality provides a novel way for charge dequantization. Specifically, we showed that a universal scale of non-locality is excluded as it would lead to the SM particles having different (dequantized) charges, which is excluded by the requirement of gauge invariance of the Yukawa couplings of fermions and the cancellation of gauge anomalies. On the other hand, we found that a flavor-dependent scale of non-locality can accommodate charge dequantization if and only if the SM neutrinos are of a Dirac type, since Majorana neutrinos imply that the electric charge of the SM is quantized. We showed that the experimental limits on the electric charge of neutrinos can set much stronger bounds on the scale of non-locality compared to the bounds from the LHC, which range between $10^{5} - 10^{10}$ TeV depending on the type of the fermion.


\section*{Acknowledgments}
The work of FA is supported by the C.V. Raman fellowship from CHEP at IISc. The work on NO is supported in part by the United States Department of Energy grant (DE-SC0012447).
\appendix
\setcounter{section}{0}
\section{More Accurate Calculation of the Uehling Potential}
\label{sec:appendix}
In Section \ref{sec:III_2}, we presented an approximate calculation to the modified electric potential (the Uehling potential). Here, we show a more accurate treatment. In momentum space, the electric potential can be expressed as
\begin{equation}\label{eq:momentum_potenial1}
	\widetilde{V}(q^{2}) = \frac{e^{2}}{q^{2}}\Big(1 - \widehat{\Pi}_{q}(q^{2})\Big),
\end{equation}
which in local QED and at 1-loop is given by
\begin{equation}\label{eq:momentum_potenial2}
	\widetilde{V}(q^{2}) = \frac{e^{2}}{q^{2}}\Big(1 + \frac{2 \alpha}{\pi}\int_{0}^{1}dxx(1-x)\log{\Big[1-x(1-x)\frac{q^{2}}{m^{2}}\Big]}\Big).
\end{equation}

Evaluating the integral and Fourier transforming the momentum space to the position space leads to the usual Uehling potential in local QED
\begin{equation}\label{eq:full_Uehling}
V(r) = -\frac{\alpha}{r}\Big(1 + \frac{\alpha}{4\sqrt{\pi}}\frac{e^{-2m r}}{(mr)^{3/2}} + \cdots\Big) , \hspace{5mm} r \gg \frac{1}{m},
\end{equation}
and we see that the first term in eq. (\ref{eq:full_Uehling}) is simply the classical Coulomb potential, whereas the second terms is a Yukawa-type interaction the arises from the quantum radiative corrections. At distances $r \gtrsim 1/m$, vacuum polarization due to virtual $e^{+}e^{-}$ pairs constitutes a screening effect that reduces the bare charge.

A similar treatment can be applied to non-local QED by using $\widehat{\Pi}_{2}(q^{2})$ in eq. (\ref{eq:reg_SE2}), where the correction to the Coulomb potential is given by
\begin{eqnarray}
\delta \widetilde{V}(q^{2}) & \simeq & \begin{cases}
\frac{8\alpha^{2}}{q^{2}} \int_{0}^{1}dx x(x-1)\text{Ei}\Big( \frac{-4m^{2}-4x(x-1)q^{2}}{\Lambda_{f}^{2}}\Big), \hspace{32mm} \text{MNS};  \label{eq:delta_V_NL}\\
\frac{8\alpha^{2}}{q^{2}} \int_{0}^{1}dx x(x-1)\Big[ \text{Ei}\Big( \frac{-4m^{2}-4x(x-1)q^{2}}{\Lambda_{f}^{2}}\Big) - \text{Ei}\Big( \frac{-4m^{2}}{\Lambda_{f}^{2}}\Big)  \Big], \hspace{5mm} \overline{\text{MNS}}.
\end{cases}
\end{eqnarray}
The integrals can be either approximated or performed numerically, then they can be Fourier transformed to yield the correction to the Coulomb potential in position space. Here we note that non-locality provides additional screening because it eliminates higher momenta.


\begin{thebibliography}{10}




\bibitem{Witten:1985cc}
E.~Witten,
``Noncommutative Geometry and String Field Theory,''
Nucl. Phys. B \textbf{268}, 253-294 (1986)

\bibitem{Kostelecky:1988ta}
V.~A.~Kostelecky and S.~Samuel,
``The Static Tachyon Potential in the Open Bosonic String Theory,''
Phys. Lett. B \textbf{207}, 169-173 (1988)

\bibitem{Kostelecky:1989nt}
V.~A.~Kostelecky and S.~Samuel,
``On a Nonperturbative Vacuum for the Open Bosonic String,''
Nucl. Phys. B \textbf{336}, 263-296 (1990)

\bibitem{Freund:1987kt}
P.~G.~O.~Freund and M.~Olson,
``NONARCHIMEDEAN STRINGS,''
Phys. Lett. B \textbf{199}, 186-190 (1987)

\bibitem{Freund:1987ck}
P.~G.~O.~Freund and E.~Witten,
``ADELIC STRING AMPLITUDES,''
Phys. Lett. B \textbf{199}, 191 (1987)

\bibitem{Brekke:1987ptq}
L.~Brekke, P.~G.~O.~Freund, M.~Olson and E.~Witten,
``Nonarchimedean String Dynamics,''
Nucl. Phys. B \textbf{302}, 365-402 (1988)

\bibitem{Frampton:1988kr}
P.~H.~Frampton and Y.~Okada,
``Effective Scalar Field Theory of $P^-$adic String,''
Phys. Rev. D \textbf{37}, 3077-3079 (1988)

\bibitem{Tseytlin:1995uq}
A.~A.~Tseytlin,
``On singularities of spherically symmetric backgrounds in string theory,''
Phys. Lett. B \textbf{363}, 223-229 (1995)
\arXivold{hep-th/9509050}.

\bibitem{Seiberg:1999vs}
N.~Seiberg and E.~Witten,
``String theory and noncommutative geometry,''
JHEP \textbf{09}, 032 (1999)
\arXivold{hep-th/9908142}.

\bibitem{Siegel:2003vt}
W.~Siegel,
``Stringy gravity at short distances,''
\arXivold{hep-th/0309093}.

\bibitem{Biswas:2004qu}
T.~Biswas, M.~Grisaru and W.~Siegel,
``Linear Regge trajectories from worldsheet lattice parton field theory,''
Nucl. Phys. B \textbf{708}, 317-344 (2005)
\arXivold{hep-th/0409089}.

\bibitem{Calcagni:2013eua}
G.~Calcagni and L.~Modesto,
``Nonlocality in string theory,''
J. Phys. A \textbf{47}, no.35, 355402 (2014)
\arXivold{1310.4957}{hep-th}.

\bibitem{Calcagni:2014vxa}
G.~Calcagni and L.~Modesto,
``Nonlocal quantum gravity and M-theory,''
Phys. Rev. D \textbf{91}, no.12, 124059 (2015)
\arXivold{1404.2137}{hep-th}.

\bibitem{Abu-Ajamieh:2023syy}
F.~Abu-Ajamieh and S.~K.~Vempati,
``A Proposed Renormalization Scheme for Non-local QFTs and Application to the Hierarchy Problem,''
\arXivold{2304.07965}{hep-th}.

\bibitem{Krasnikov:1987yj}
N.~V.~Krasnikov,
``NONLOCAL GAUGE THEORIES,''
Theor. Math. Phys. \textbf{73}, 1184-1190 (1987)

\bibitem{Moffat:1988zt}
J.~W.~Moffat,
``FINITE QUANTUM FIELD THEORY BASED ON SUPERSPIN FIELDS,''
Phys. Rev. D \textbf{39}, 3654 (1989)

\bibitem{Moffat:1990jj}
J.~W.~Moffat,
``Finite nonlocal gauge field theory,''
Phys. Rev. D \textbf{41}, 1177-1184 (1990)

\bibitem{Biswas:2005qr}
T.~Biswas, A.~Mazumdar and W.~Siegel,
``Bouncing universes in string-inspired gravity,''
JCAP \textbf{03}, 009 (2006)
\arXivold{hep-th/0508194}.

\bibitem{Biswas:2014yia}
T.~Biswas and N.~Okada,
``Towards LHC physics with nonlocal Standard Model,''
Nucl. Phys. B \textbf{898}, 113-131 (2015)
\arXivold{1407.3331}{hep-ph}.


\bibitem{Capolupo:2022awe}
A.~Capolupo, G.~Lambiase and A.~Quaranta,
``Muon $g-2$ anomaly and non-locality,''
Phys. Lett. B \textbf{829}, 137128 (2022)
\arXivold{2206.06037}{hep-ph}.

\bibitem{Capolupo:2023kuu}
A.~Capolupo, A.~Quaranta and R.~Serao,
``Phenomenological implications of nonlocal quantum electrodynamics,''
\arXivold{2305.17992}{hep-ph}.

\bibitem{Briscese:2015zfa}
F.~Briscese, E.~R.~Bezerra de Mello, A.~Y.~Petrov and V.~B.~Bezerra,
``One-loop effective potential in nonlocal scalar field models,''
Phys. Rev. D \textbf{92}, no.10, 104026 (2015)
\arXivold{1508.02001}{gr-qc}.


\bibitem{Aoyama:2020ynm}
T.~Aoyama, N.~Asmussen, M.~Benayoun, J.~Bijnens, T.~Blum, M.~Bruno, I.~Caprini, C.~M.~Carloni Calame, M.~C\`e and G.~Colangelo, \textit{et al.}
``The anomalous magnetic moment of the muon in the Standard Model,''
Phys. Rept. \textbf{887}, 1-166 (2020)
\arXivold{2006.04822}{hep-ph}.

\bibitem{Muong-2:2006rrc}
G.~W.~Bennett \textit{et al.} [Muon g-2],
``Final Report of the Muon E821 Anomalous Magnetic Moment Measurement at BNL,''
Phys. Rev. D \textbf{73}, 072003 (2006)
\arXivold{hep-ex/0602035}{hep-ex}.

\bibitem{Muong-2:2021ojo}
B.~Abi \textit{et al.} [Muon g-2],
``Measurement of the Positive Muon Anomalous Magnetic Moment to 0.46 ppm,''
Phys. Rev. Lett. \textbf{126}, no.14, 141801 (2021)
\arXivold{2104.03281}{hep-ex}.

\bibitem{Muong-2:2021ovs}
T.~Albahri \textit{et al.} [Muon g-2],
``Magnetic-field measurement and analysis for the Muon $g-2$ Experiment at Fermilab,''
Phys. Rev. A \textbf{103}, no.4, 042208 (2021)
\arXivold{2104.03201}{hep-ex}.

\bibitem{Muong-2:2021vma}
T.~Albahri \textit{et al.} [Muon g-2],
``Measurement of the anomalous precession frequency of the muon in the Fermilab Muon $g-2$ Experiment,''
Phys. Rev. D \textbf{103}, no.7, 072002 (2021)
\arXivold{2104.03247}{hep-ex}.

\bibitem{Muong-2:2023cdq}
D.~P.~Aguillard \textit{et al.} [Muon g-2],
``Measurement of the Positive Muon Anomalous Magnetic Moment to 0.20 ppm,''
\arXivold{2308.06230}{hep-ex}.

\bibitem{Borsanyi:2020mff}
S.~Borsanyi, Z.~Fodor, J.~N.~Guenther, C.~Hoelbling, S.~D.~Katz, L.~Lellouch, T.~Lippert, K.~Miura, L.~Parato and K.~K.~Szabo, \textit{et al.}
``Leading hadronic contribution to the muon magnetic moment from lattice QCD,''
Nature \textbf{593}, no.7857, 51-55 (2021)
\arXivold{2002.12347}{hep-lat}.

\bibitem{Ce:2022kxy}
M.~C\`e, A.~G\'erardin, G.~von Hippel, R.~J.~Hudspith, S.~Kuberski, H.~B.~Meyer, K.~Miura, D.~Mohler, K.~Ottnad and P.~Srijit, \textit{et al.}
``Window observable for the hadronic vacuum polarization contribution to the muon g-2 from lattice QCD,''
Phys. Rev. D \textbf{106}, no.11, 114502 (2022)
\arXivold{2206.06582}{hep-lat}.

\bibitem{ExtendedTwistedMass:2022jpw}
C.~Alexandrou \textit{et al.} [Extended Twisted Mass],
``Lattice calculation of the short and intermediate time-distance hadronic vacuum polarization contributions to the muon magnetic moment using twisted-mass fermions,''
Phys. Rev. D \textbf{107}, no.7, 074506 (2023)
\arXivold{2206.15084}{hep-lat}.


\bibitem{Abu-Ajamieh:2022nmt}
F.~Abu-Ajamieh and S.~K.~Vempati,
``Can the Higgs Still Account for the g-2 Anomaly?,''
\arXivold{2209.10898}{hep-ph}.

\bibitem{Abu-Ajamieh:2023qvh}
F.~Abu-Ajamieh, M.~Frasca and S.~K.~Vempati,
``Flavor Violating Di- Higgs Coupling,''
\arXivold{2305.17362}{hep-ph}.

\bibitem{Buttazzo:2020ibd}
D.~Buttazzo and P.~Paradisi,
``Probing the muon $g-2$ anomaly with the Higgs boson at a muon collider,''
Phys. Rev. D \textbf{104}, no.7, 075021 (2021)
\arXivold{2012.02769}{hep-ph}.

\bibitem{Yin:2020afe}
W.~Yin and M.~Yamaguchi,
``Muon g-2 at a multi-TeV muon collider,''
Phys. Rev. D \textbf{106}, no.3, 033007 (2022)
\arXivold{2012.03928}{hep-ph}.

\bibitem{Fajfer:2021cxa}
S.~Fajfer, J.~F.~Kamenik and M.~Tammaro,
``Interplay of New Physics effects in $(g- 2)_{l}$ and $h \rightarrow l^{+}l^{-}$ lessons from SMEFT,''
JHEP \textbf{06}, 099 (2021)
\arXivold{2103.10859}{hep-ph}.

\bibitem{Aebischer:2021uvt}
J.~Aebischer, W.~Dekens, E.~E.~Jenkins, A.~V.~Manohar, D.~Sengupta and P.~Stoffer,
``Effective field theory interpretation of lepton magnetic and electric dipole moments,''
JHEP \textbf{07}, 107 (2021)
\arXivold{2102.08954}{hep-ph}.

\bibitem{Allwicher:2021jkr}
L.~Allwicher, L.~Di Luzio, M.~Fedele, F.~Mescia and M.~Nardecchia,
``What is the scale of new physics behind the muon g-2?,''
Phys. Rev. D \textbf{104}, no.5, 055035 (2021)
\arXivold{2105.13981}{hep-ph}.

\bibitem{Cheung:2021iev}
K.~Cheung and Z.~S.~Wang,
``Physics potential of a muon-proton collider,''
Phys. Rev. D \textbf{103}, 116009 (2021)
\arXivold{2101.10476}{hep-ph}.

\bibitem{Abu-Ajamieh:2023roj}
F.~Abu-Ajamieh, P.~Chattopadhyay, A.~Ghoshal and N.~Okada,
``Anomalies in String-inspired Non-local Extensions of QED,''
\arXivold{2307.01589}{hep-th}.

\bibitem{Eides:2000xc}
M.~I.~Eides, H.~Grotch and V.~A.~Shelyuto,
``Theory of light hydrogen - like atoms,''
Phys. Rept. \textbf{342}, 63-261 (2001)
\arXivold{hep-ph/0002158}.

\bibitem{Matveev:2013orb}
A.~Matveev, C.~G.~Parthey, K.~Predehl, J.~Alnis, A.~Beyer, R.~Holzwarth, T.~Udem, T.~Wilken, N.~Kolachevsky and M.~Abgrall, \textit{et al.}
``Precision Measurement of the Hydrogen 1S\ensuremath{-}2S Frequency via a 920-km Fiber Link,''
Phys. Rev. Lett. \textbf{110}, no.23, 230801 (2013)

\bibitem{Babu:1989tq}
K.~S.~Babu and R.~N.~Mohapatra,
``Is There a Connection Between Quantization of Electric Charge and a Majorana Neutrino?,''
Phys. Rev. Lett. \textbf{63}, 938 (1989)

\bibitem{Babu:1989ex}
K.~S.~Babu and R.~N.~Mohapatra,
``Quantization of Electric Charge From Anomaly Constraints and a Majorana Neutrino,''
Phys. Rev. D \textbf{41}, 271 (1990)

\bibitem{Foot:1992ui}
R.~Foot, H.~Lew and R.~R.~Volkas,
``Electric charge quantization,''
J. Phys. G \textbf{19}, 361-372 (1993)
[erratum: J. Phys. G \textbf{19}, 1067 (1993)]
\arXivold{hep-ph/9209259}.

\bibitem{Das:2020egb}
A.~Das, D.~Ghosh, C.~Giunti and A.~Thalapillil,
``Neutrino charge constraints from scattering to the weak gravity conjecture to neutron stars,''
Phys. Rev. D \textbf{102}, no.11, 115009 (2020)
\arXivold{2005.12304}{hep-ph}.

\bibitem{Giunti:2014ixa}
C.~Giunti and A.~Studenikin,
``Neutrino electromagnetic interactions: a window to new physics,''
Rev. Mod. Phys. \textbf{87}, 531 (2015)
\arXivold{1403.6344}{hep-ph}.

\bibitem{Raffelt:1999gv}
G.~G.~Raffelt,
``Limits on neutrino electromagnetic properties: An update,''
Phys. Rept. \textbf{320}, 319-327 (1999)


\end{thebibliography}
\end{document}